\documentclass[12pt]{article}
\usepackage{epsf}

\newcommand{\Tint}[1]{{\hbox{$\sum$}\!\!\!\!\!\!\int}_{\!\!\!\!#1}}
\newcommand{\la}[1]{\label{#1}}
\newcommand{\beq}{\begin{equation}}
\newcommand{\eeq}{\end{equation}}
\newcommand{\bea}{\begin{eqnarray}}
\newcommand{\eea}{\end{eqnarray}}
\newcommand{\be}{\begin{equation}}
\newcommand{\ee}{\end{equation}}
\newcommand{\ba}{\begin{eqnarray}}
\newcommand{\ea}{\end{eqnarray}}
\newcommand{\bi}{\begin{itemize}}
\newcommand{\ei}{\end{itemize}}
\newcommand{\rmi}[1]{{\mbox{\scriptsize #1}}}
\newcommand{\fig}{Fig.~}
\newcommand{\eq}{Eq.~}
\newcommand{\eqs}{Eqs.~}
\newcommand{\nr}[1]{(\ref{#1})}
\newcommand{\tr}{{\rm Tr\,}}
\newcommand{\nn}{\nonumber}
\newcommand{\fr}[2]{{\frac{#1}{#2}}}
\newcommand{\msbar}{\overline{\mbox{\rm MS}}}
\newcommand{\lambdamsbar}{\Lambda_{\overline{\rm MS}}}

\newcommand{\bfx}{{\bf x}}

\newcommand{\bmu}{\bar{\mu}}
\newcommand{\se}{Sec.~}
\renewcommand{\vec}[1]{{\bf #1}}
\def\mb#1         {\mbox{\boldmath $#1$}}

\def\lsi{\raise0.3ex\hbox{$<$\kern-0.75em\raise-1.1ex\hbox{$\sim$}}}
\def\gsi{\raise0.3ex\hbox{$>$\kern-0.75em\raise-1.1ex\hbox{$\sim$}}}
\newcommand{\lsim}{\mathop{\lsi}}
\newcommand{\gsim}{\mathop{\gsi}}

\newcommand{\re}{{\rm Re}}
\newcommand{\im}{{\rm Im}}
\newcommand{\ii}{{\rm i}}
\makeatletter \@addtoreset{equation}{section} \makeatother
\renewcommand{\theequation}{\arabic{section}.\arabic{equation}}

\begin{document}

\begin{titlepage}
\begin{flushright}
Edinburgh 2000/09\\
CERN-TH/2000-105\\
hep-ph/0004060
\end{flushright} 
\begin{centering} 

\vfill

{\bf STATIC CORRELATION LENGTHS IN QCD AT\\
HIGH TEMPERATURES AND FINITE DENSITIES}
\vspace{0.8cm}

A. Hart$^{\rm a}$,
M. Laine$^{\rm b,c}$ and 
O. Philipsen$^{\rm b}$

\vspace{0.3cm}
{\em $^{\rm a}$%
Dept.\ of Physics and Astronomy, Univ.\ of Edinburgh,\\
Edinburgh EH9 3JZ, Scotland, UK\\}
\vspace{0.3cm}
{\em $^{\rm b}$%
Theory Division, CERN, CH-1211 Geneva 23,
Switzerland\\}
\vspace{0.3cm}
{\em $^{\rm c}$%
Dept.\ of Physics, P.O.Box 9, FIN-00014 Univ.\ of Helsinki,
Finland\\}

\vspace{0.7cm}

{\bf Abstract}

\end{centering}

\vspace{0.3cm}
\noindent
We use a perturbatively derived effective field theory and 
three-dimensional lattice simulations to determine the longest 
static correlation lengths in the deconfined QCD plasma phase at high 
temperatures ($T\gsim 2 T_c$) and finite densities ($\mu \lsim 4 T$). 
For vanishing chemical potential, we refine a previous determination
of the Debye screening length, and determine the dependence of different 
correlation lengths on the number of massless flavours as well as 
on the number of colours. For non-vanishing 
but small chemical potential, the existence of Debye screening 
allows us to carry out simulations corresponding to the full QCD with 
two (or three) massless dynamical flavours, in spite of a complex action. 
We investigate how the correlation lengths in the different quantum number 
channels change as the chemical potential is switched on. 


\vfill

\noindent
Edinburgh 2000/09\\
CERN-TH/2000-105\\
May 2000

\vfill

\end{titlepage}

\section{Introduction}

With the advent of RHIC and ALICE, there is a growing need for a
precise understanding of various properties of QCD at temperatures up
to a GeV. At the moment, we are still far from a satisfactory level in
this respect, even for the equilibrium properties of the plasma.
Indeed, even though the system can in principle be described as a gas
of quarks and gluons, a fully perturbative computation with these
degrees of freedom does in practice not work well at any reasonable
temperatures below $\sim 10^{10}$ GeV, since the perturbative series
is badly convergent due to infrared sensitive
contributions~\cite{gpy}--\cite{ad}.  On the other hand, the only
systematic fully non-perturbative method available, four-dimensional
(4d) lattice simulations, is severely restricted in the presence of
light dynamical fermions, and even more so in the presence of a finite
baryon density; the current state of the art is summarised
in~\cite{lat99}. A first principles solution of any problem related to
non-equilibrium phenomena remains even further in the future.

The physics problem we address in this paper is the determination
of various static correlation lengths. To tackle this problem, 
we will employ a method which can overcome 
some of the difficulties mentioned above.  
The method combines analytic and numerical
techniques, in a way that both are used in a regime where they are well
manageable. First, perturbation theory is employed in deriving, by 
what is called dimensional reduction~\cite{dr,rold,hl,generic,bn,ad}, 
an effective theory for the long range degrees of freedom of the system.
At high temperatures all such degrees of freedom are bosonic, since
fermions are screened by non-zero Matsubara frequencies $\pi T n$,
with $n$ odd. The reduction step can thus be carried out with
dynamical, massless fermions and, as we shall see, with a finite
chemical potential. Second, non-perturbative lattice simulations are
used to study the infrared sensitive dynamics of the remaining degrees
of freedom. The theory to be studied with simulations is the 
SU(3)+adjoint Higgs model in three dimensions (3d), and many of its
properties have already been determined~\cite{su3adj}.

The method we employ suffers clearly from a number of restrictions.
First of all, in QCD it is limited to temperatures above $T_c$,
roughly $T \gsim 2 T_c$~\footnote{For the electroweak theory, in
  contrast, similar methods allow for a precise determination of the
  properties of the phase transition at $T=T_c$ for Higgs masses
  $m_H\sim 30...250$ GeV \cite{nonpert,cfh}.}. This can be inferred
from a comparison of the dynamical scales described by the effective
theory to those integrated out~\cite{ad} as well as, more concretely,
from a direct quantitative comparison of the correlation lengths
measured within 3d for SU(2)~\cite{hp}, with those determined using 4d
SU(2) lattice simulations at finite temperature (without
fermions)~\cite{dg}\footnote{For early work with similar conclusions
  both in SU(2) and SU(3), see~\cite{rold}. Recently, analogous
  results have been reached also by considering
  gauge fixed correlators~\cite{fop}, as well as by considering
  dimensional reduction of pure SU(3) from (2+1)d to 2d~\cite{rnew}.}.  
Second, as we shall see, it
is also limited by the largest chemical potential that can be reached.
We will be able to go up to $\mu\lsim 4 T$.

If there are restrictions, there are also strengths. What can 
be done within the effective theory can be done quite precisely. 
The simulations are comparatively easy technically, they yield high 
accuracy continuum limits for the quantities under investigation, 
and hence there is little ambiguity in the conclusions.
For instance, studies in 3d have produced detailed and accurate
insight into the 
relative sizes of perturbative and non-perturbative contributions to the
Debye screening length \cite{lp,hp}, a quantity which 
could be directly relevant for such signals of the 
quark--gluon plasma as $J/\Psi$ suppression. 

In the present work, we extend the study of \cite{hp} first to 
SU(3) Yang-Mills theory, and then to two- and three-flavour QCD in the 
massless limit. In comparing with earlier results in SU(3)~\cite{ad,lp},
we improve on the accuracy of Debye screening length determination
and measure the screening lengths also in the other quantum number
channels. We also discuss explicitly the $N_c$ and $N_f$
dependence of our results --- here $N_c$ is the number of colours
and $N_f$ is the number of (massless) quark flavours.  

We then proceed to extend our calculations to finite baryon density.
Our approach is straightforward: dimensional reduction leads to an
additional complex term in the effective action, which has to be
absorbed in the observables for practical simulations.  This
``reweighting'' displays in principle similar problems as chemical
potential simulations in four dimensions (for reviews,
see~\cite{lat99,alf}).  We find that for the effective theory there exists
a range of volumes and ratios $\mu/T$, however, for which the problems
are in practice manageable. We demonstrate this by investigating a
range of imaginary chemical potentials~\cite{imagmu}, for which we
find complete agreement between reweighted calculations and those
using the exact action. We then employ the reweighting technique to
the case of real chemical potentials.

Let us note that our approach 
relies on $\mu/(\pi T)$ being small, but no restriction is imposed
on fermionic masses $m_i, i=1,...,N_f$, 
which we take to be zero. In 4d simulations, 
on the other hand, progress has been possible when the $m_i$ are
taken to be very large~\cite{bigm,ippl}, whereas the temperature 
can be small.

\section{Continuum formulation for $\mu=0$}
\la{contform}

We start by reviewing the result of the dimensional 
reduction step for $\mu=0$. We follow closely Ref.~\cite{ad}.

The effective theory emerging from hot QCD by dimensional reduction \cite{dr}
is the SU(3)+adjoint Higgs model with the action
\begin{equation} \label{actc}
        S = \int d^{3}x \left\{ \frac{1}{2} \tr F_{ij}^2
        +\tr [D_{i},A_0]^2 +m_3^{2} \tr A_0^2 
        +\lambda_3(\tr A_0^2)^{2} \right\} ,
\end{equation}
where $F_{ij}=\partial_{i}A_{j}-\partial_{j}A_{i}
 +ig[A_i,A_j]$,
$D_i = \partial_i + ig_3 A_i$, 
$F_{ij},A_i$, and $A_0$ are all traceless $3\times 3$
Hermitian matrices ($A_0=A_0^{a}T_{a}$, etc), and $g_3^2$ and $\lambda_3$
are the gauge and scalar coupling constants with mass dimension one, 
respectively.
The physical properties of the effective theory are determined by
the two dimensionless ratios 
\beq
x=\frac{\lambda_3}{g_3^2}, \quad y=\frac{m_3^2(\bmu_3=g_3^2)}{g_3^4},
\la{xy}
\eeq
where $\bmu_3$ is the $\msbar$ dimensional regularization
scale in 3d. The parameters $x,y$, as well as the scale $g_3^2$, are
via dimensional reduction functions of  the temperature $T$
and the QCD scale $\lambdamsbar$. For $N_c=3$, the result is \cite{ad}
\bea \label{3d1}
g_3^2&=&\frac{24\pi^2}{33-2N_f}\;\frac{1}{\ln(\bmu_g/\lambdamsbar)}
\;T,\\
x&=&\frac{9-N_f}{33-2N_f}\;\frac{1}{\ln(\bmu_x/\lambdamsbar)},
\la{3d2}\\
y(x)&=&\frac{(9-N_f)(6+N_f)}{144\pi^2 x}+
\frac{486-33N_f-11N_f^2-2N_f^3}{96(9-N_f)\pi^2} + {\cal O}(x). \la{3d3}
\eea
We have assumed here $N_f$ massless flavours, but the
inclusion of quark masses is also possible in principle. 

The scales entering in \eqs\nr{3d1}, \nr{3d2} are 
\ba \label{muopt}
\bmu_i & = & \bmu_T\hat{\mu}_i, \\
\bmu_T & = & 4\pi e^{-\gamma_E}T\approx 7.0555T, \qquad  
\hat{\mu}_i = 
\exp\biggl({-3c_i+4N_f\ln4\over66-4N_f}
\biggr), \\
c_g & =  & 1,\qquad
c_x = {54-22N_f\over 9-N_f}+\fr43 N_f. \la{cg}
\ea
To be explicit, for $N_f=2$ we obtain
\be
g_3^2 = \frac{24\pi^2 T}{29\ln(8.11\, T/\lambdamsbar)}, \quad
x = \frac{7}{29\ln(6.91\, T/\lambdamsbar )}, \quad
y = \frac{7}{18\pi^2 x} + \frac{15}{28\pi^2}.  \la{nf2}
\ee
Corrections to these expressions are of relative magnitude
$\sim {\cal O}(\alpha_s/\pi)^2\sim {\cal O}(x^2)$.

An important question is how small one can
in practice take $T/\lambdamsbar$ and
still have only small corrections in \eqs\nr{3d1}--\nr{3d3}. As was
observed in~\cite{ad}, for $y(x)$ the expansion appears to converge 
surprisingly fast, with an error on a few percent level even at 
$T\sim \lambdamsbar$. The expansions need {\em a priori} not 
be as good for every parameter, however. 

Consider for instance $g_3^2$. One may expect higher loop 
graphs to amount to an increase in the effective scale factor $\bmu_g$, 
since there are many ``massive'' ($\sim 2\pi T$)
particles inside the loops instead of one. Thus an upper bound 
on the error in $g_3^2$ can be obtained by evaluating the 2-loop QCD
running coupling at the 1-loop scale $\bmu_g\sim 7 T$
(for $N_f=0$). This gives a 
correction of up to 25\%, which is quite large.
An explicit computation of $g_3^2$ including 
effects of order ${\cal O}(g^6)$ would thus be welcome in order to 
clarify whether the expansion for $g_3^2$ in reality converges as rapidly as
for $y(x)$ or not. Such computations are beyond our scope 
here and we will simply use \eqs\nr{3d1}--\nr{3d3}; a comparison
with direct 4d simulations at $N_f=0$ turns out to be quite
satisfactory within this procedure, and thus we will work under the 
assumption that the error in $g_3^2,x$ is similarly small
as in~$y$.

Finally, we may sometimes want to express the temperature
to which our simulations correspond in terms of $T_c$, 
rather than $\lambdamsbar$. In view of \eq\nr{nf2},  
this requires knowledge of the deconfinement temperature $T_c/\lambdamsbar$,
which can only be fixed with 4d simulations.
For the case $N_f=0$, we take the value $T_c/\lambdamsbar=1.03(19)$ 
from~\cite{4dpt}\footnote{A recent computation~\cite{bfp} favours
a slightly larger central value, but well within the error bars.}. 
For $N_f\neq 0$, the determination of $T_c$ is not
accurate, and we will take $T_c/\lambdamsbar=1.0$ as a reference.
Our results however only depend on $T/\lambdamsbar$, so they can be 
reinterpreted as corresponding to some other temperature
relative to $T_c$ if need be. 

\subsection{Operators and quantum numbers}
\la{qn}

The physical observables on which we shall focus in this paper are
spatial correlation lengths of QCD at finite temperatures.  
For the practical calculations we
reinterpret the 3d theory in \eq\nr{actc} which we use to compute them, 
as a (2+1)d one.  We take
operators to live on the ($x_1,x_2$)-plane, and the correlations are
taken in the $x_3$ direction. Thus we compute the spectrum of the 2d
Hamiltonian of the effective theory, whose eigenvalues then correspond
to {\it screening masses} or inverse spatial correlation lengths of
the 4d theory at finite temperature.  Thus, unless otherwise stated,
``bound states'', ``glueballs'' etc. refer to the eigenstates of the
effective Hamiltonian and to the screening masses of the 4d theory in
the above sense.

We use the quantum number notation 
\ba
R: & & D_i \to D_i,\quad F_{ij} \to F_{ij}, \quad A_0 \to -A_0, \la{defR}\\
P: & & D_1 \to D_1,\quad D_2\to -D_2,
\quad F_{12}\to -F_{12},\quad A_0\to A_0, \\
C: & & D_i\to D_i^*,\quad F_{ij}\to -F_{ij}^*, \quad  A_0 \to -A_0^*. \la{defC}
\ea
The action in \eq\nr{actc} is invariant under these operations. 
In the finite temperature context, the operation $R$ in the 3d theory
is a remnant of the 4d time reversal operation~\cite{ay}.  The parity
$P$ means a parity on the 2d plane, i.e.\ a reflection across the
$x_1$-axis.  The charge conjugation $C$ is a non-trivial quantum
number only for SU(3), since for SU(2) it is just a global gauge
transformation $i\tau^2$, so that there are no gauge invariant
operators odd in $C$~\cite{ay}. In addition to these discrete transformations,
we define a rotation in the ($x_1,x_2$)-plane, with the corresponding
angular momentum $J$. Thus the full symmetry group is $SO(2)\otimes
Z_2(R) \otimes Z_2(P)\otimes Z_2(C)$, and accordingly we classify our
operators and states by $J^{PC}_R$.

We note from \eqs\nr{defR}--\nr{defC}
that apart from $D_2$, the action of $RPC$
on the operators in the ($x_1,x_2$)-plane corresponds to complex
conjugation (and $x_2\to -x_2$). 
Thus any real operator which does not contain 
$D_2$ is even in $RPC$. This means, in particular, that
for the scalar $J=0$ operators in which no single $D_2$ can appear,
there are effectively only two 
quantum numbers left out of the original three, thus four
different channels. 
The lowest dimensional gauge invariant operators
of these types in the $J=0$ channels are: 
\ba
J^{PC}_R= 0^{++}_{+}: & & \tr A_0^2, \tr F_{12}^2, ... \nn\\
J^{PC}_R= 0^{--}_{+}: & & \tr F_{12}^3, \tr A_0^2F_{12}, ... \nn\\
J^{PC}_R= 0^{-+}_{-}: & & \tr A_0 F_{12}, ... \nn\\
J^{PC}_R= 0^{+-}_{-}: & & \tr A_0^3, \tr A_0 F_{12}^2, ...\la{contop}
\ea
All operators here with $C=-1$ vanish identically for SU(2). 
Going out of the plane (i.e., allowing also for $F_{13},F_{23}$), 
other channels would in principle become possible~\cite{ay}, 
but we do not expect them to change our conclusions.

\section{Lattice formulation for $\mu=0$}
\la{contaction}

To simulate the effective theory
in \eq\nr{actc} on the lattice, we use the discretised version
\begin{eqnarray}
S &=& \beta \sum_{\bfx,i > j}\left(1-\frac{1}{3}\re
     \tr U_{ij}(\bfx)\right)
     +2\sum_{\bfx} \tr(\varphi(\bfx)\varphi(\bfx)) \nonumber \\
  &-&2\kappa \sum_{\bfx,i}\tr(\varphi(\bfx)U_{i}
      (\bfx)\varphi(\bfx+{\hat{\imath}})U^{\dagger}_i(\bfx))
      +\lambda \sum_{\bfx}(2 \tr(\varphi(\bfx)\varphi(\bfx))-1)^{2},
      \label{lattice_action}
\end{eqnarray}
where the fields in the lattice action have been rescaled relative
to the continuum, 
$A_0(\bfx)= ({\kappa/a})^{1/2} \varphi(\bfx)$
where $a$ is the lattice spacing, and $U_{ij}(\bfx)$
denotes the elementary $1\times 1$ plaquette in the $i,j$-plane 
located at $\bfx$.
The parameters of the continuum and lattice theory are up to two loops
related by a set of equations which become exact in the continuum 
limit~\cite{framework,contlatt}, 
\bea
\beta&=&\frac{6}{ag_3^2}, \quad
\lambda=\frac{3 x\kappa^2}{2 \beta}, \nn\\
y&=& \frac{\beta^2}{18}
                \left(\frac{1}{\kappa}-3
                -\frac{3x\kappa}{\beta}\right)
                +\frac{3.1759115\,\beta}{4\pi}
                     \left(1+\frac{5}{3}x\right) \nonumber\\
                &+&\frac{1}{16\pi^{2}}
                  \left[(60x-20x^2)
                  \left(\ln\beta+0.08849\right)
                  +34.768x+36.130\right].
\eea
For a given pair of continuum parameters
$x,y$ these equations determine the lattice parameters
$\kappa,\lambda$ as a function of the lattice spacing, and hence govern
the approach to the continuum limit, $\beta\rightarrow \infty$.
We have not implemented ${\cal O}(a)$ improvement~\cite{moore_a}
here, since the discretisation effects 
in the correlation lengths are already quite small 
at the values of $\beta$ we are using.

\subsection{Observables}

Operators for all the quantum number channels considered can be constructed
from a number of basic operator types.
We consider operators involving only scalar fields or products of scalar
and gauge field variables,
\bea
R_2(\bfx)&=&\tr(\varphi^2(\bfx)), \nn\\
R_3(\bfx)&=&\tr(\varphi^3(\bfx)), \nn\\
L_i(\bfx)&=&\tr \left( \varphi(\bfx)U_i(\bfx)
\varphi(\bfx+{{\hat{\imath}}})U_i^{\dag}(\bfx) \right), \nn\\
B_1(\bfx)&=&\tr(\varphi(\bfx) U_{ij}(\bfx)), \nn\\
B_2(\bfx)&=&\tr(\varphi^2(\bfx) U_{ij}(\bfx)).
\eea
Furthermore, we have loop operators constructed from link variables only,
\beq
 C_{ij}^{1\times1}(\bfx)=\tr\left(
 U_i(\bfx)U_j(\bfx+{{\hat{\imath}}})
 U^\dag_i(\bfx+{{\hat{\jmath}}})U^\dag_j(\bfx) \right),
 \quad i,j=1,2,\,i\not=j,
\eeq
and in addition to the elementary plaquette $C^{1\times1}$, we also
consider squares of size $2\times2$ as well as rectangles of size
$1\times2$, $1\times3$, $2\times3$. 
Another useful pure gauge
operator is the Polyakov loop along a {\em spatial} direction $j$,
\beq
  P_j^{(L)}(\bfx)=\re\tr\prod_{m=0}^{L-1}\,
  U_j(\bfx+m{{\hat{\jmath}}}), \quad j=1,2,
\eeq
which can be used to extract the 3d string tension. It also 
gives useful information about finite volume effects via so called
torelon states~\cite{mic,ptw}.

Operators with definite quantum number assignments are constructed
from the above types by taking linear combinations with
appropriate transformation properties. Clearly operators
containing an even number of scalar fields, such as $B_2$, will
couple to $R = +1$ states, and those with odd, such as $B_1$,
to $R = -1$. The different $P,C$ channels can be chosen by 
utilising the projection operators $\fr12(1\pm P)$, $\fr12(1\pm C)$ 
where, e.g., $C B_1(x_1,x_2) = -B_1^*(x_1,x_2)$, 
$P B_1(x_1,x_2) = B_1^*(x_1,-x_2)$. 
Different spin states are obtained by employing
the operation $R(\theta_n)$ which rotates the operators
by $n (\pi/2)$ around~$\bfx$. Note also that even though 
lattice rotations are restricted to multiples of $\pi/2$, 
we are in practice close enough to the continuum limit that
continuum symmetries are reproduced within the statistical errors;
thus we keep the continuum notation in terms of $J$ even on the lattice.

In the following we list the operators used 
corresponding to \eq\nr{contop}: 
\begin{tabbing}
\indent \= $P_d$:\,\, \= \kill
$0^{++}_+$ channel: \\
\> $R$:   \> $R_2(\bfx)$, \\
\> $L$:   \> $\re \left(L_1(\bfx)+L_2(\bfx)\right)$, \\
\> $C$:   \> real part of symmetric combinations of 
             $C^{1\times1},\,C^{2\times2},\,
             C^{1\times2},\,C^{1\times3},\,C^{2\times3}$, \\
\> $P$:   \> $P_1^{(L)}(\bfx)+P_2^{(L)}(\bfx)$, \\
\> $T$: \> $\left(P_1^{(L)}(\bfx)\right)^2+\left(P_2^{(L)}(\bfx)\right)^2$,\\
\> \\
$0^{--}_+$ channel: \\
\> $C$:   \> imaginary part of symmetric combinations of
             $C^{1\times1},\,C^{2\times2},\,
             C^{1\times2},\,C^{1\times3},\,C^{2\times3}$,\\
\> $B$:   \> $\im \sum_{n=1}^4 R(\theta_n) B_2(\bfx)$, \\
\> \\
$0^{-+}_-$ channel: \\
\> $B$:   \> $\im \sum_{n=1}^4 R(\theta_n) B_1(\bfx)$, \\
\> \\
$0^{+-}_-$ channel: \\
\> $R$:   \> $R_3(\bfx)$, \\
\> $B$:   \> $\re \sum_{n=1}^4 R(\theta_n) B_1(\bfx)$. \\
\end{tabbing}
We have also measured higher spin states.  For $2^{++}_+$ we have a
fairly large basis, identical to the one described in~\cite{hp}. 
The $J=1$ states prove to be quite heavy, so that their relevance
for the 4d finite temperature system is not 
immediately obvious at modest temperatures, and thus
for simplicity we consider here only the ground states in the 
channels with $R=\pm 1$, without specifying the other quantum numbers.

\subsection{Blocking and matrix correlators}

For later reference, let us recall the general principles
of how the operators discussed above can be used for a reliable
extraction of correlation lengths~\cite{gev}. 

The eigenstates of the 2d Hamiltonian in the region of
parameter space where we work are bound states. 
In order to increase the overlap of our operators 
onto such extended states, we construct smeared 
link and scalar field variables $\phi_i$ as described in~\cite{hp}.
For every quantum number channel, we then measure the correlation matrix
\beq
C_{ij}(t)=\langle \phi_i^\dagger(t)\phi_j(0)\rangle, \la{cij}
\eeq
where we have denoted $t=x_3$.
This matrix can be diagonalised by solving a generalised eigenvalue problem,
\beq \label{evp}
C^{-1}(0)C(t) \vec{v}_n =\lambda_n(t) \vec{v}_n,
\eeq
where $\lambda_n(t) \sim \exp(-a M_n t)$. We carry 
out this procedure at $t=a,2a,3a,$ which gives somewhat
differing eigenvectors $\vec{v}_n$, and check that the 
final outcome remains the same within error bars. We normalise
the eigenvectors according to $\vec{v}_n^\dagger C(0) \vec{v}_n = 1$, 
so that $\Phi_n = \sum_i \vec{v}_n^{(i)} \phi_i$ 
satisfy $\Phi_n^\dagger \Phi_n = 1$, and 
are thus the normalized eigenstates of the 2d Hamiltonian. 
The final results are extracted from the correlation functions 
\beq \label{Mt}
G_n(t)=\langle \Phi_n^\dagger(t) \Phi_n(0)\rangle=
\vec{v}_n^\dagger C(t)\vec{v}_n,
\eeq
by computing effective masses
\beq
a M_\rmi{eff,{\em n}}(t)=-\ln\left[\frac{G_n(t+1)}
{G_n(t)}\right] \, ,
\eeq
and ensuring that these have attained a stable plateau value.
Information about the composition of $\Phi_n$ in terms
of the operators used in the simulation, which we shall 
quote in some of our tables, is obtained from the overlaps 
$\langle \phi_i^\dagger \Phi_n \rangle = 
\sum_j C_{ij}(0)\vec{v}_n^{(j)}$. 

\subsection{Simulation and analysis}

In our Monte Carlo simulation of the lattice action, 
\eq (\ref{lattice_action}),
link variables are updated by a combination of heat bath and over-relaxation
steps with algorithms described in \cite{teper99}. The scalar fields are
generated by a combination of heat bath and reflection steps~\cite{bunk}.
One ``compound'' sweep consists of several over-relaxation and reflection
updates following each heat bath update of gauge and scalar fields.
Measurements are taken after every compound sweep. Typically, we gathered
between 5 000 and 20 000 measurements depending on the lattice sizes.
Statistical errors are estimated using a jackknife procedure with bin
sizes of 100 -- 250 measurements.

\section{Numerical results for $\mu=0$}
\la{numres}

In this section we discuss simulations of the effective theory for
parameter values corresponding to hot SU(3) gauge theory with
$N_f=0,2,3,4$ flavours of fermions at zero baryon density.
Detailed investigations of finite volume effects and scaling
of correlation lengths have been
performed for the 3d pure SU(2) and SU(2)+adjoint Higgs theories 
in \cite{teper99,hp}, as well as for the 3d pure SU(3) theory
in \cite{teper99,lp}. In these works 
explicit continuum extrapolations were performed. 
For large enough $\beta$ the scaling behaviour was found to be quite
good, with continuum results differing from those on a fine lattice by
only a few percent. Our gauge couplings here were chosen to produce a
similar lattice spacing, and checks also show good scaling.
The parameter combinations 
and the lattices used for them are collected 
in Table \ref{tab_params}.
\begin{table}[htb]
\begin{center}
\begin{tabular}{|c|r@{.}l|r@{.}l|r@{.}l|l|l|}
\hline
\hline
$N_f$ & \multicolumn{2}{c|}{$T/T_c$} &
\multicolumn{2}{c|}{$x$} & \multicolumn{2}{c|}{$y$} &
$\beta = 21$ & $\beta = 28$ \\
\hline
$0$  & 2&0 & 0&11346 & 0&39188 &
        $L^2 \cdot T = 28^3$, $38^3$ & $L^2 \cdot T = 40^3$ \\
\mbox{}& \multicolumn{2}{c|}{$\sim 10^{11}$} & 0&009636 & 4&0 &
        --- & $L^2 \cdot T = 40^3$  \\
$2$  & 1&5 & 0&10304 & 0&43668 &
        $L^2 \cdot T = 30^3$ & --- \\
\mbox{}& 2&0 & 0&09191 & 0&4830 & 
        $L^2 \cdot T = 30^3$ & $L^2 \cdot T = 40^3$ \\
$3$  & 1&5 & 0&08680 & 0&47890 &
        $L^2 \cdot T = 30^3$ & --- \\
\mbox{}& 2&0 & 0&07814 & 0&52741 &
        $L^2 \cdot T = 30^3$ & --- \\
$4$  & 2&0 & 0&062921 & 0&569680 &
        $L^2 \cdot T = 30^3$ & --- \\
\hline
\hline
\end{tabular}
\caption{ \label{tab_params}
{\em The lattice parameters and sizes used for calculations at $\mu=0$.
Let us stress that for $N_f>0$, $T/T_c$ means really 
$T/\lambdamsbar$, see Sec.~\ref{contform}.}}
\end{center}
\end{table}

\subsection{The mass spectrum for $N_f=0$} 

The main features of the spectrum, displayed in detail in 
Tables \ref{tab_02_0nf}, \ref{tab_other_0nf}
in the Appendix, are the same as those found
in previous studies of the confinement region for SU(2) with fundamental
\cite{ptw} or adjoint \cite{hp} scalar fields. There is a dense spectrum 
of bound states, consisting of a replication of the glueball spectrum
found in the $d=2+1$ pure SU(3) theory \cite{teper99}, and additional
bound states of scalars, with little mixing between the two. 
This may be concluded from the fact that, as indicated in 
Table \ref{tab_02_0nf},
some eigenstates are composed predominantly of purely gluonic
operators $C$, with practically no contribution from operators 
carrying the same quantum numbers but containing scalars. Comparing with
\cite{teper99}, we find that our 
gluonic states have quantitatively the same masses as the
corresponding glueballs in the pure Yang-Mills theory; the
comparison is shown in Table \ref{tab_comp_pg}.

\begin{table}[tb]
\begin{center}
\begin{tabular}{|l|r@{.}lr@{.}l|r@{.}l|}
\hline
 & \multicolumn{4}{|c|}{gauge--Higgs} & \multicolumn{2}{c|}{pure gauge} \\
\cline{2-7}
$J_R^{PC}$ & \multicolumn{2}{|c}{scalar} & \multicolumn{2}{c|}{glueball} &
\multicolumn{2}{c|}{glueball} \\
\hline
\hline
$0^{++}_+$ &
0&994 (19)  &  2&511 (65)     &  2&575 (18)  \\
 & 
2&95 (16)   &   3&61 (32)     &  3&841 (28)    \\
\hline
\hline
$0^{--}_+$  &
2&46 (10)   &  3&50 (25)     &  3&795 (27)  \\
\hline
\hline
$2^{++}_+$ &
3&355 (94)  &  4&14 (28)     &  4&257 (33)  \\
\hline
\end{tabular}
\caption{ \label{tab_comp_pg}
  {\em Mass estimates $M/g_3^2$ at $T=2T_c$ and
    $N_f=0$. 
    Our data for the glueballs are compared with the results
    obtained in the pure gauge theory \cite{teper99}.}}
\end{center}
\end{table}

As can be observed from Table~\ref{tab_comp_pg}, however,
in the physically interesting region of couplings the lightest
states in given quantum number channels are
not glueballs but always states including
the scalar field $\varphi$. This can be understood in the sense
that in the finite temperature context 
the phase transition is (for $N_f=0$) assumed to be driven
by the $Z(N)$ symmetry related to $A_0$ (or $\varphi$
on the lattice), and thus
$\varphi$ is the dominant infrared degree of freedom. 
Let us now discuss the corresponding correlation 
lengths in more detail. 

\subsubsection{The ``magnetic'' sector, $R=+1$}
\la{magsec}

\begin{figure}[tb]%

\vspace*{-3.5cm}

\begin{center}\leavevmode
\epsfysize=250pt
\epsfbox[70 80 570 680]{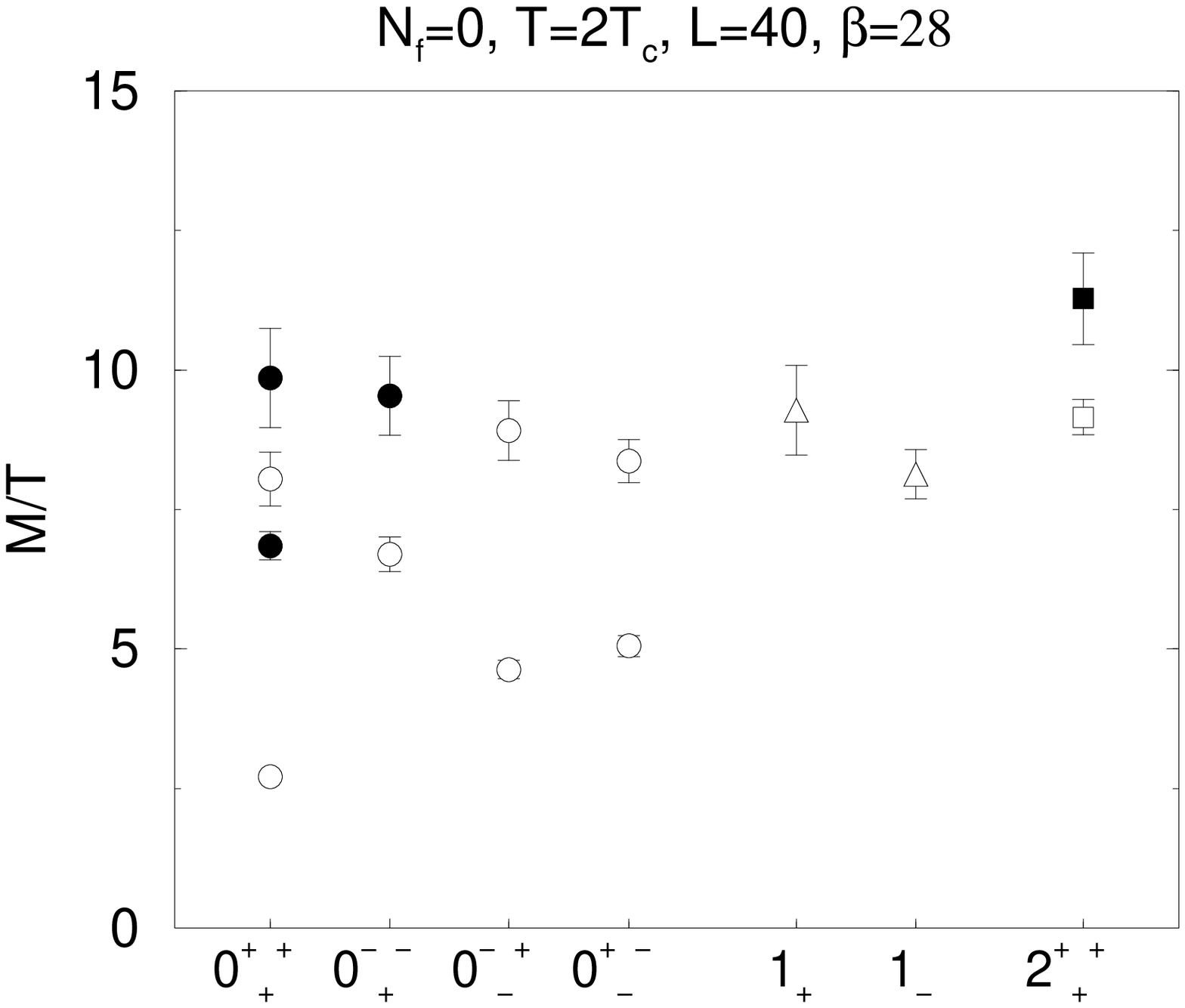}
\leavevmode
\epsfysize=250pt
\epsfbox[70 80 570 680]{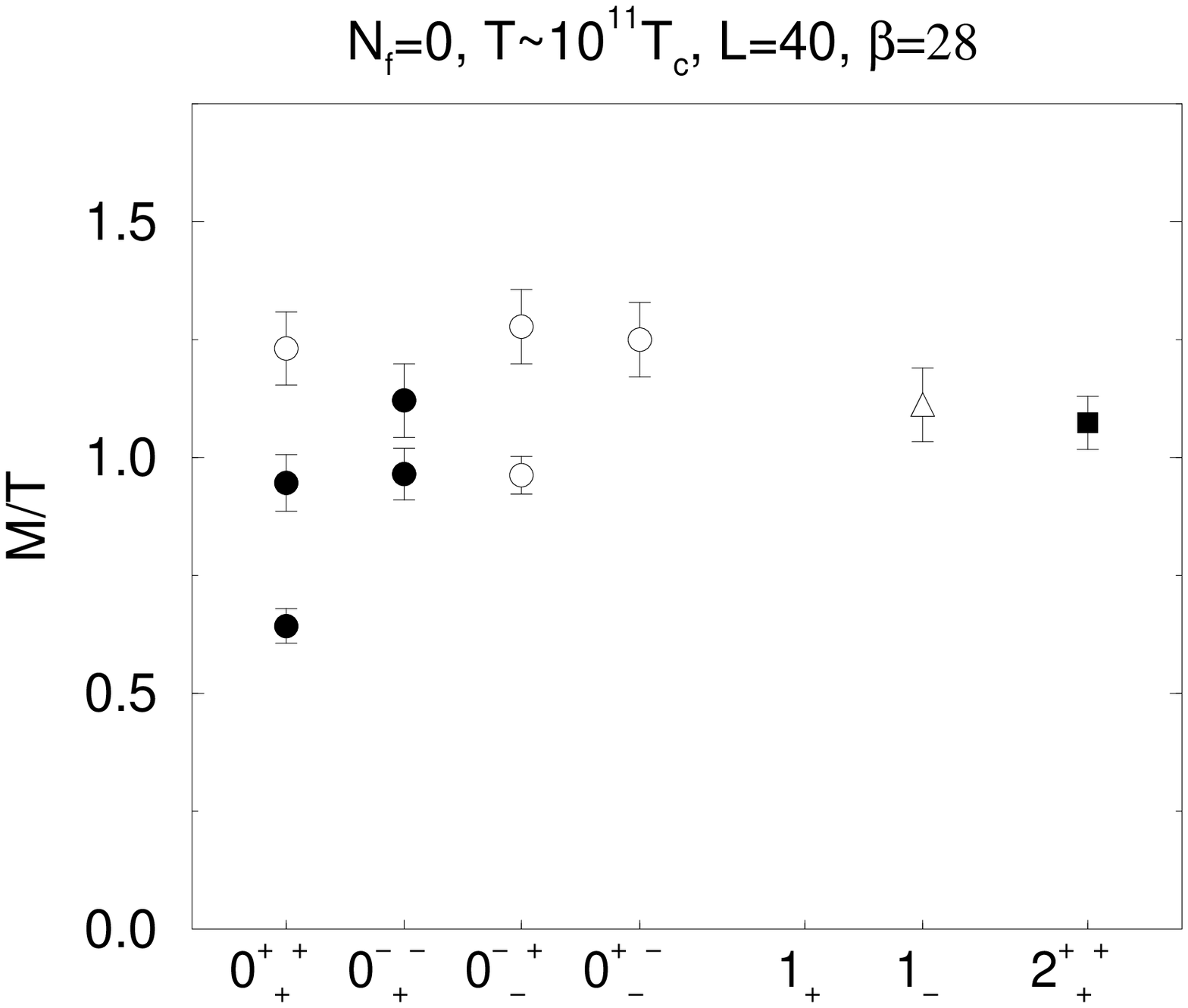}
\end{center}


\caption[]{\label{spec_nf0}{\it
The spectrum of screening masses in various quantum number channels
at $N_f=0, T=2T_c$ (left), $N_f=0, T\sim 10^{11} T_c$ (right).
Filled symbols denote 3d glueball states, which have become the 
lightest excitations at $ T\sim 10^{11} T_c$. The states $1_+$ 
are much heavier than $1_-$ 
at high temperatures, and thus not visible on the right.}}
\end{figure}

Consider first operators with $R=+1$.
We will call this the ``magnetic'' sector, even though the 
operators can include an even number of $A_0$-fields. This sector
determines the correlation length related to the lowest lying
glueballs and all other $0^{++}_+$ observables, as well as that 
felt by the real part of the 4d temporal Polyakov loop~\cite{pol,ay}
(not to be confused with the {\it spatial} Polyakov loop in the 3d
theory). In order to express the masses in units of temperature, we use 
the perturbative expression for the gauge coupling in \eq\nr{3d1}
to arrive at $g_3^2(T=2T_c)\approx 2.7 T$ and 
$g_3^2(T=10^{11}T_c)\approx 0.25 T$.  The spectrum in these 
units is shown in \fig \ref{spec_nf0}. 

Let us first discuss the channel $0_{+}^{++}$.
{}From Table~\ref{tab_02_0nf} in the Appendix, we note
that the lightest state (open circle in \fig\ref{spec_nf0}(left))
is dominantly $\sim \tr A_0^2$, while the next state  
(filled circle in \fig\ref{spec_nf0}(left)) is dominantly 
$\sim \tr F_{12}^2$. Thus, the na\"{\i}ve 
ordering of states consisting of the fields $A_0\sim gT$ and $A_i\sim g^2T$
is reversed non-perturbatively: at low temperatures it is the $A_0$ which
is responsible for the lightest gauge invariant state in the system.

As the temperature is increased, however, the ordering gets changed.
The lowest state becomes heavier with increasing temperature, 
as the scalar ``constituent mass parameter'' ${y}^{1/2}$ is then growing
(for a discussion of the constituent picture in a similar 
context, see~\cite{bp2,fop}). On the other hand, 
the glueball $\sim\tr F_{12}^2$ 
does not contain much $A_0$ admixture. Correspondingly 
its mass, in units of $g_3^2$, is quite insensitive to $y$,  
so that for large enough temperatures 
indeed the glueball corresponds to the 
lightest excitation. This situation is, however, only realised 
at high temperatures $T > 10^2 T_c$; an 
extreme example is shown in \fig \ref{spec_nf0}(right). 
In this limit the $A_0$ may be integrated out,
leaving the 3d pure SU(3) gauge
theory as an effective theory~\cite{dr}.

For the channel $0_+^{--}$, the behaviour is similar, but
the masses are much larger. Let us mention here already that 
in the 4d theory with fermions at modest temperatures $T\sim 2T_c$, 
all the higher-lying bosonic states in general are probably 
heavier than some gauge invariant states 
consisting of fermionic fields $\sim \bar\psi \psi$, which are not 
addressed by our theory, and whose mass we may expect to be $\lsim 2\pi T$.
(At lower temperatures such states can 
be even lighter, since in the chiral limit 
they are expected to represent the critical degrees of freedom.)

To get an impression of the quality of dimensional reduction for $N_f=0$, 
we now compare the $0_+^{++}$ state $\sim \tr A_0^2$ with that
found directly from a 4d simulation at $N_f=0$~\cite{dg}.
Our value is $M[0^{++}_+]=2.71(6)T$, to be compared
with $M=2.60(4)T$ reported in \cite{dg}.
In complete analogy with the SU(2) case~\cite{hp}, it is thus found that
dimensional reduction quantitatively reproduces the lowest
screening mass of the 4d Yang-Mills theory at temperatures as low
as $T=2T_c$. 

We can also carry out another comparison. The real part of the 4d temporal
Polyakov loop carries quantum numbers $0^{++}_+$, and 
non-perturbatively mixes with other operators in that channel. 
Correspondingly,
its decay should also be determined
by the bound state $\tr A_0^2$. Indeed, the exponential decay of the
4d temporal Polyakov loop correlator is in 4d 
measured to be $\sim 2.5 T$~\cite{fk},
which is quite compatible with the value 
$\approx 2.7 T$ we find here.

On the other hand, in contrast to SU(2) where even several excitations
agree between the full and the effective theory, we find here
deviations of about 20\% in comparing with~\cite{dg}.  This may signal
that the effective theory is not as reliable for states whose mass is
approaching $\sim 2\pi T$.  However, 
we also believe that there could be some
room for improvement upon this apparent disagreement.  The 4d
simulations are quite difficult, and it is not easy to guarantee
at this stage that the infinite volume and continuum limits are
reached for the excited states.

\subsubsection{The ``electric'' Debye sector, $R=-1$}
\label{debye}

Of particular phenomenological relevance for the QCD plasma is the Debye 
mass, whose inverse gives the length scale over which the colour-electric
field is screened. The expansion in powers
of coupling constants for this quantity reads
\be
\frac{m_D}{g_3^2}=\frac{m_D^{LO}}{g_3^2}+{N_c\over4\pi}\ln{m_D^{LO}\over g_3^2}
+c_{N_c} + \frac{{\cal O}(g^3T)}{g_3^2}. \la{mDexp}
\ee 
The perturbative leading order result is \cite{lo}
\beq \label{dm}
m_D^{LO}=\left(\frac{N_c}{3}+\frac{N_f}{6}\right)^{1/2}gT
= g_3^2\, \sqrt{y} + {\cal O}(g^3 T),
\eeq
but at next-to-leading order $\sim g_3^2$, only a logarithm can 
be extracted~\cite{rebhan} (the 2nd term on the right hand side
of \eq\nr{mDexp}), 
whereas the coefficient $c_{N_c}$ is entirely
non-perturbative. To determine it, 
we use the gauge invariant definition of the Debye mass 
based on the Euclidean time reflection symmetry as given in \cite{ay}.
According to this definition, the Debye mass corresponds to the mass
of the lightest state odd under this transformation. The remnant of Euclidean
time reflection symmetry in our reduced model is the scalar reflection
symmetry $R$, and the Debye mass thus corresponds to the
mass of the lightest $R=-1$ state of our spectrum.
As shown in \fig\ref{spec_nf0}, this is the
$0^{-+}_-$ ground state $\sim \tr A_0 F_{12}$, 
while $0^{+-}_-$ $\sim \tr A_0^3$ is slightly heavier. 

With this definition, the coefficient $c_{N_c}$ can 
be measured separately from the exponential 
decay of a Wilson line in a 3d pure gauge theory~\cite{ay}. 
The measurements for $N_c=2,3$ have 
been performed in \cite{lp} with the results $c_2=1.14(4)$, $c_3=1.65(6)$.
On the other hand, the mass we have measured here
also includes the ${\cal O}(g^3T)$ correction in \eq\nr{mDexp}. 
We can thus now estimate the magnitude of the remainder;
our results are shown in Table~\ref{deb}. We find that the 
${\cal O}(g^3T)$ corrections 
are less than 30\% even at temperatures as low as $T=2T_c$, 
and they disappear entirely for asymptotically large temperatures. 
The sign is negative, so that the complete result is 
somewhat smaller than estimated in~\cite{lp} based on 
the perturbative contributions and $c_3$ alone, 
$\sim 5T$ instead of $\sim 6T$. The result is also 
about $20$\% smaller than what can be extracted from~\cite{ad}, 
a difference which we presume to be due to the relatively 
small lattice sizes used there. 
\begin{table}
\begin{center}
\begin{tabular}{|l|r@{.}l|r@{.}l|r@{.}l|r@{.}l|r@{.}l|}
\hline
\hline
 &\multicolumn{2}{|c|}{$m_D/g_3^2$} &
\multicolumn{2}{|c|}{pert.\ part} &
\multicolumn{2}{|c|}{$c_3$} &
\multicolumn{2}{|c|}{${\cal O}(g^3T)/g_3^2$} &
\multicolumn{2}{|c|}{$m_D/T$} \\
\hline
$T=2T_c$            & 1&70(5)  & 0&514 & 1&65(6) & -0&46(6)  & 4&6(2)  \\
$T\sim 10^{11}T_c$  & 3&82(12) & 2&165 & 1&65(6) &  0&00(12) & 0&96(3) \\
\hline
\hline\end{tabular}
\caption{ \label{deb}
{\em The different contributions to the Debye mass, \eq\nr{mDexp}.
The perturbative part means the 1st and 2nd terms on the 
right-hand-side of \eq\nr{mDexp}.}} 
\end{center}
\end{table}
 
We can conclude that for temperatures of physical interest,
the Debye mass according to this definition is entirely non-perturbative,
with its next-to-leading order correction being larger than the leading
term. This remains to be the case up to $T\sim 10^7 T_c$ \cite{lp}. But 
even at $T\sim 10^{11}T_c$, as the table shows, there are sizeable corrections
to the leading behaviour. 
This leads to the conclusion that the scale dominating
the Debye mass is $\sim g^2 T$ with non-perturbative physics 
for all temperatures of interest, 
in contrast to the na\"{\i}ve expectation $\sim gT$. The reason for this
behaviour is the large non-perturbative coefficient $c_{N_c}$ that appears
in front of the correction term in the series in $g$.

\subsubsection{$N_c$ scaling between SU(2) and SU(3)}

\begin{figure}[tb]%

\vspace*{-3.5cm}

\begin{center}\leavevmode
\epsfysize=250pt
\epsfbox[70 80 570 680]{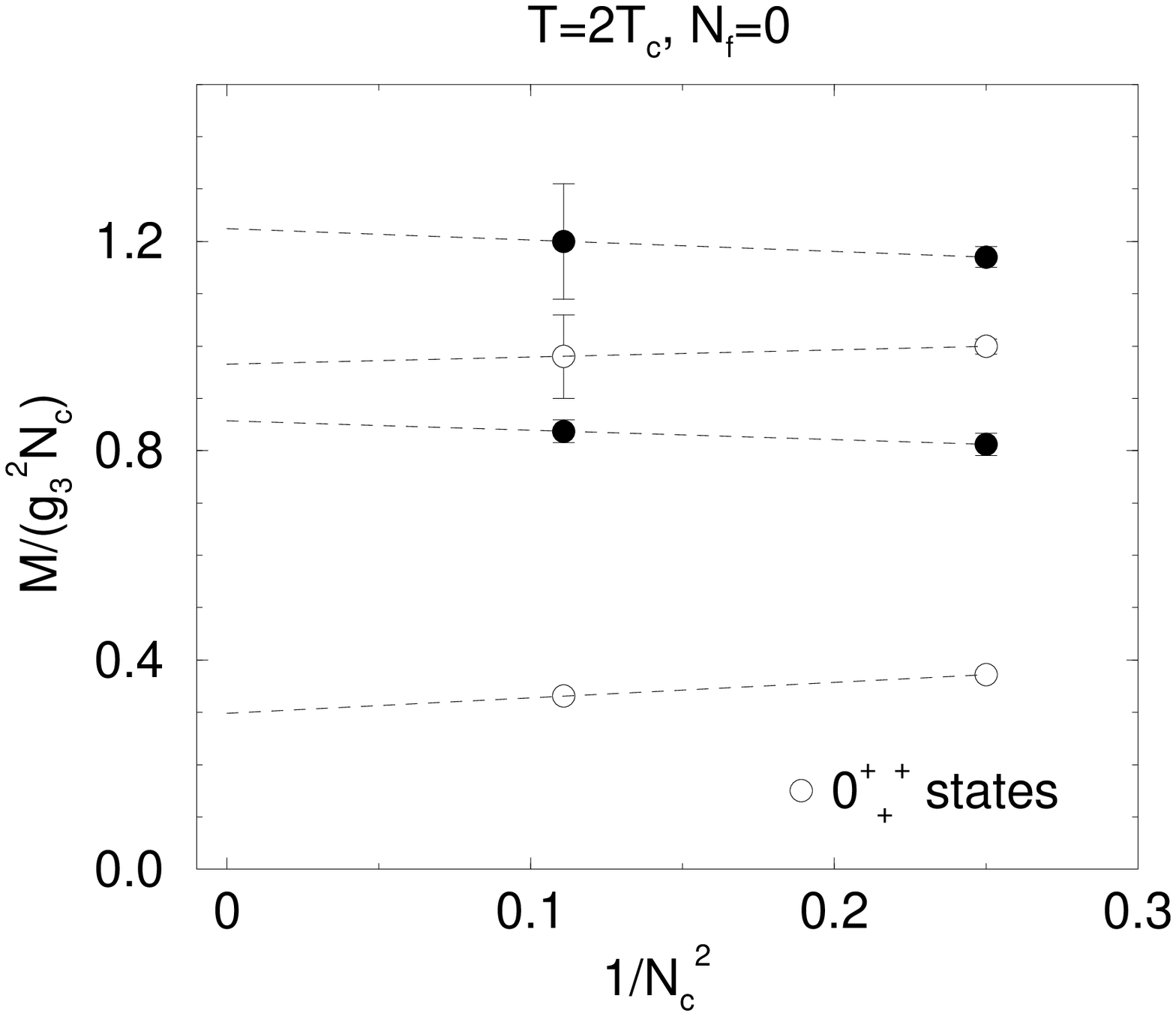}
\leavevmode
\epsfysize=250pt
\epsfbox[70 80 570 680]{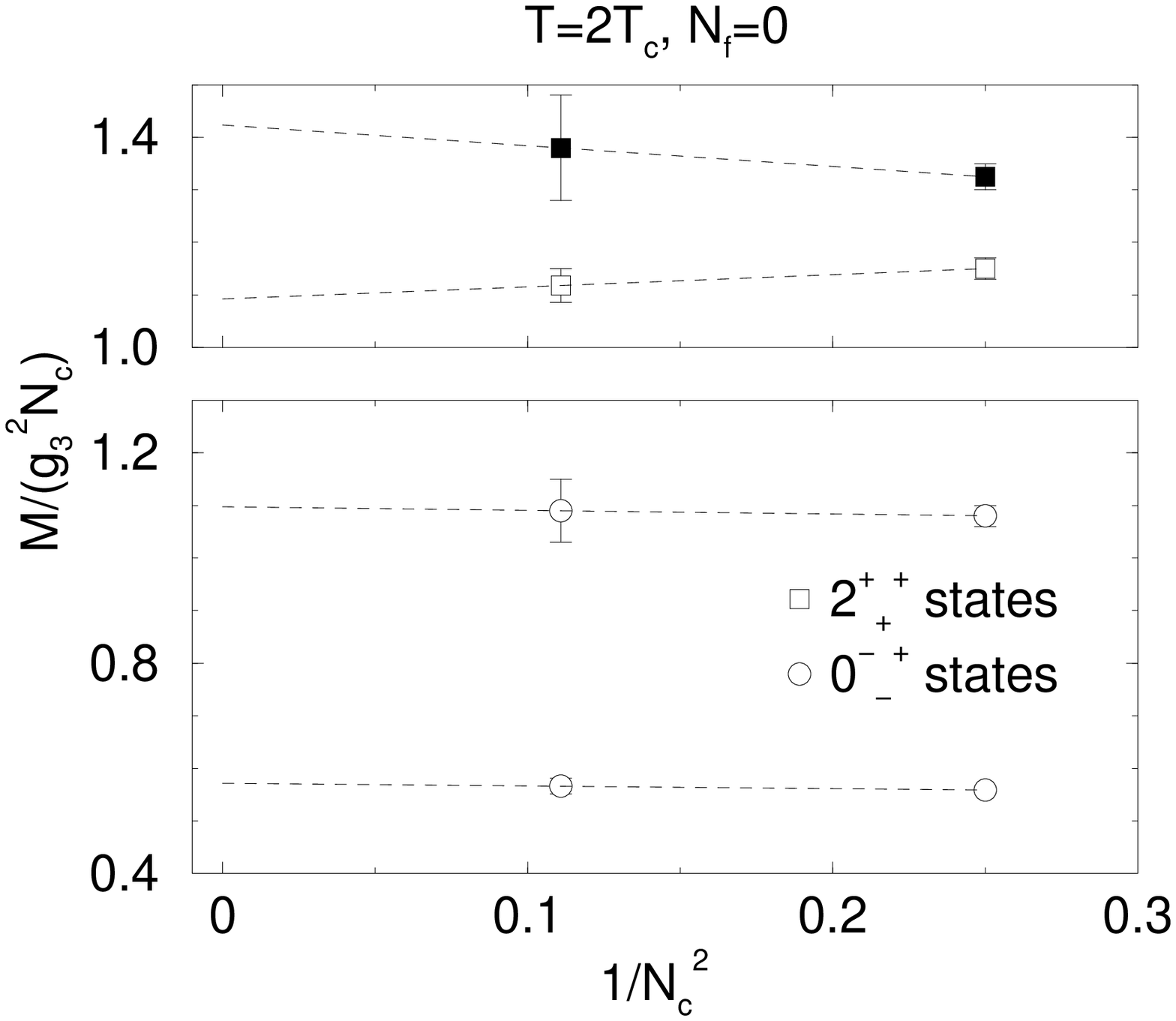} 
\end{center}


\caption[]{\label{nc}{\it
The scaling with $N_c$ of some of the low lying $0_+^{++}$ (left) and 
$0_-^{-+}, 2_+^{++}$ (right) states.
Filled symbols denote glueballs, and the lines are to guide the eye only.}}
\end{figure}

Let us finally discuss the scaling with $N_c$. In \cite{teper99} it
was found for SU($N_c$) pure gauge theories with $N_c=2,...5$ that the
differences in the mass spectra can be accounted for by the leading
order $1/N_c^2$ corrections in a large $N_c$ expansion, 
and that the coefficients of these are
remarkably small.  In theories with various scalar fields, the
glueball content has been found to be practically identical to that of
pure gauge theories for $N_c=2,3$~\cite{ptw,iss,hp}, and thus the
scaling behaviour with $N_c$ is preserved. It is then natural to ask
if the same scaling behaviour holds for the scalar bound states, which
are not present in the Yang-Mills theory. In \fig \ref{nc} we plot
some of our low lying states from the current analysis as well as the
SU(2) case~\cite{hp} (we have to focus on states with $C=+1$ in order
to have an SU(2) counterpart). Indeed we observe a similar scaling for
the scalar states as for the glueballs. This suggests that the 
screening masses of hot SU(3) gauge theory are
close to the $N_c \to \infty$ limit,
and hence large $N_c$ methods may
be useful approximations to analytically deal with some of the
non-perturbative aspects discussed here.

\subsection{The mass spectrum for $N_f\neq 0$}

Having convinced ourselves that dimensional reduction of pure Yang-Mills
theory works at the very least for the lowest static bosonic
correlation functions
of the theory, we may now move ahead to exploit the advantage of this
formalism, the easy inclusion of fermions. For the case of massless fermion,
all that is required is to evaluate \eqs\nr{3d1}--\nr{3d3} 
for the desired $N_f$
and simulate the effective theory with the corresponding parameters.
Each of these simulations gives results completely analogous to those
of the $N_f=0$ case, except for a slight shift in the values.
For the reader interested in the detailed numbers, 
we collect our data in Tables~\ref{tab_02}, \ref{tab_other} in the Appendix, 
at temperatures $T=2\lambdamsbar$, 
and $1.5\lambdamsbar$, which probes the lower limit
of the range of validity of dimensional reduction.

\begin{figure}[tb]%

\vspace*{-5.0cm}

\begin{center}\leavevmode
\epsfysize=350pt
\epsfbox[20 30 620 730]{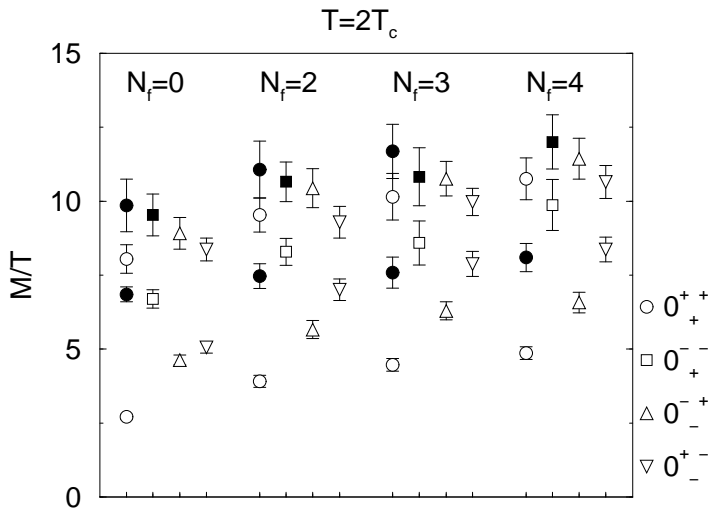}
\end{center}

\vspace{-1.0cm}

\caption[]{\label{spec_nf}{\it
The $N_f$ dependence of the $J=0$ spectrum at $T = 2T_c$
(to be more precise, for $N_f>0$  $T_c$ means really 
$\lambdamsbar$, see Sec.~\ref{contform}).
Filled symbols denote glueball states. We should stress that 
only the part $M \lsim 2 \pi T$ of the spectrum can be expected
to directly represent the lowest states in 4d finite temperature QCD.}}
\end{figure}

With the tables at hand, the spectrum shown in \fig \ref{spec_nf0} for $N_f=0$
is now also known for $N_f=2,3,4$. In \fig \ref{spec_nf} we
display the dependence on $N_f$ of the low lying $J=0$ states.
The general behaviour of slightly rising mass values with $N_f$ can be 
understood from \eqs\nr{mDexp}, \nr{dm}. 
Increasing $N_f$ increases the value of $m_D^\rmi{LO}$ 
corresponding to the bare scalar mass, and hence the masses of the 
bound states increase as well. In units of the temperature, 
the increase from $N_f=0$ to the phenomenologically interesting
case $N_f=2...3$ is about 30\% for $0_+^{++}$, 
$\lsim 20$\% for the other channels. 
The correlation lengths decrease accordingly. 

\subsection{The spatial string tension}

Finally, let us discuss one observable of the 3d theory which 
does not have an interpretation as a physical 4d correlation length. 
A Polyakov loop 
in a {\em spatial} direction couples
to a flux loop state (or torelon)
that winds around the periodic boundaries of the finite volume.
The exponential fall-off of its correlator
is related to the mass of the flux loop, according to
\beq
\label{string}
  \sum_{\vec{x},j=1,2}\,\left\langle P_j^{(L)}(\vec{x}+t\,{\hat{3}})
   P_j^{(L)}(\vec{x})
                  \right\rangle \simeq e^{-aM_P(L)t},\quad
   aM_P(L)=a^2\sigma_L L.
\eeq
Such a flux loop state can be easily identified through its energy
scaling linearly with the size of the lattice as seen in, e.g.,
Table~\ref{tab_sigma_0nf}. Since the scalar fields are in the adjoint
representation, they cannot screen the colour flux represented by the
Polyakov loop in fundamental representation, in contrast to the case
with fundamental scalars \cite{sb}. Thus, a string tension can be
defined by the slope of the static potential at infinite separation,
just as in pure gauge theory. Accordingly, the coefficient $\sigma_L$
corresponds to the string tension at separation $L$. For large enough
$L$, an estimate for the string tension in infinite volume is then
provided by the relation
\cite{for85}
\beq
\label{stringinfty}
  a^2\sigma_\infty = a^2\sigma_L+\frac{\pi}{6L^2}.
\eeq
We have extracted the string tension by diagonalising a basis of 
various smeared Polyakov loops,
and the results are given in Tables 
\ref{tab_sigma_0nf}, \ref{tab_sigma}.

We note that the string tension in continuum units is rather insensitive
to the precise parameter values $x,y$, as it fluctuates by less than
$\sim$ 4\% between the cases considered. Moreover, all values are
similarly close to the ones obtained in pure gauge theory \cite{teper99}. 
This is a further indication of the insensitivity of the pure gauge sector 
to the presence of the adjoint scalar fields.
In the context of Yang-Mills theory at finite temperature, 
the string tension of the
effective theory corresponds to the spatial string tension as measured
in 4d simulations at finite temperature. Indeed, our value
$\sqrt{\sigma_{\infty}}/g_3^2=0.569(10)$ on the finest lattice 
for $T=2T_c$, $N_f=0$ is in good agreement with the spatial 4d
result $\sqrt{\sigma_s}/g^2T=0.586(5)$ given in \cite{kll}.

\section{Extension of the method to finite density}

As we have mentioned in the Introduction, there is no satisfactory algorithm
to simulate lattice QCD in four dimensions at finite baryon density,
with small quark masses. On the other hand, in the experimental 
situation of heavy ion collisions there always is a net baryon 
density, which in the collisions at and above AGS and SPS energies
can be estimated to correspond to $\mu/T \lsim 4.0$~\cite{exp}. 
As long as the temperature is sufficiently
above $T_c$, dimensional reduction is applicable 
for the lowest lying static correlation
functions in QCD, as we have seen in the previous
sections. We now proceed to dimensionally reduce QCD at finite 
temperature {\it and} density. As we shall see, it is possible to
perform simulations in the $\mu/T$-range of interest in this framework.

\subsection{The effective theory with $\mu\neq 0$}

At leading order, the introduction of $\mu\neq 0$ leads to
a very simple modification of the effective 3d theory
in \eq\nr{actc}. New effects come only from fermions, 
where we change $p_f\to p_f - i \mu$ in the loop momenta;
here $p_f$ denotes the fermionic Matsubara frequencies. 
The 1-loop effective potential computed in~\cite{kaps}
tells then that the effective action for the
Matsubara zero mode of $A_0$ has
for $\mu\neq 0$ the terms ($N_c=2,3$)
\beq
V_{A_0}=
g^2 \biggl[T^2\Bigl(\frac{N_c}{3} + \frac{N_f}{6}\Bigr) +
\mu^2 \frac{N_f}{2\pi^2} \biggr] \tr A_0^2 +
i g^3\mu \frac{N_f}{3\pi^2} \tr A_0^3 +
g^4 \frac{6+N_c-N_f}{24\pi^2} \Bigl( \tr A_0^2 \Bigr)^2. \la{1lS}
\eeq
There is of course no term linear
in $A_0$, unlike in the Abelian case~\cite{ks96}. 
Note furthermore that for SU(2), $\tr A_0^3=0$.

In general, the inclusion of $\mu\neq 0$ also leads to 
other new operators than $\tr A_0^3$. 
However, as can be seen from the effective
potential computed in~\cite{kaps}, there are no higher order 
operators involving only $A_0$, at least at 1-loop and 2-loop levels.
On the contrary, there could be operators such as 
$\sim \tr A_0 F_{ij}^2$. As is usual in effective field 
theories, we expect such non-renormalizable operators
to give contributions suppressed with respect 
to the dynamical effects arising within the effective
theory in \eq\nr{1lS} by a power of the scale hierarchy, 
and therefore we ignore them here. 
It is perhaps also worthwhile to mention that since
no $\gamma_5$ appears in perturbation theory in QCD
and since the construction of the effective theory
in \eq\nr{1lS} is only sensitive to the ultraviolet and 
thus purely perturbative, no Chern-Simons type operators
are expected to be generated. 

Going to 3d units ($A_0^\rmi{4d} \to T^{1/2} A_0^\rmi{3d}$, 
$\int\! d\tau\,d^3 x\to T^{-1} \int\! d^3x$) and
denoting
\beq
S_z = \int \! d^3 x\, g_3^3 \tr A_0^3,  \la{sz}
\eeq
the dominant changes due to $\mu\neq 0$ 
in the action in \eq\nr{actc} are then
\beq
S \to S + i z S_z, \quad
z= \frac{\mu}{T} \frac{N_f}{3\pi^2}; \quad
y\to y\biggl(1+\Bigl( \frac{\mu}{\pi T}\Bigr)^2 \frac{3 N_f}{2 N_c + N_f}
\biggr). \la{lo-formulas}
\eeq
Thus, one new operator is generated in the effective action, 
and one of the parameters which already existed, gets modified.
We note that the new operator is quite 
special and it, for instance, does not generate new ultraviolet
divergences for the parameters in the original action
in \eq\nr{actc}.

\begin{figure}[tb]

\centerline{
\epsfxsize=6.5cm\epsfbox{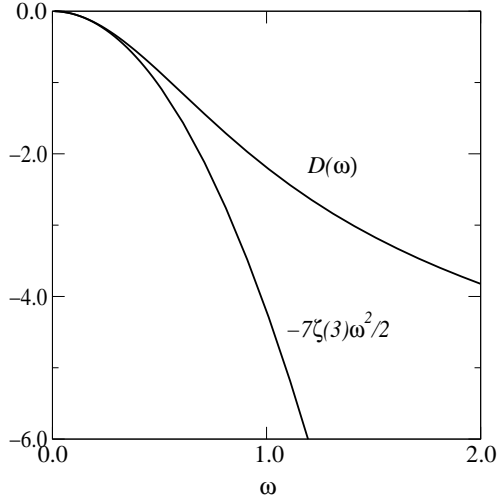}}

\caption[a]{The function ${\cal D}(\omega)$ from \eq\nr{Dx},
compared with the small-$\omega$ limit.}
\la{fig:Dx}
\end{figure}

A more precise computation
of the effective action requires the inclusion of 2-loop
effects in \eq\nr{1lS}, as well as a 1-loop computation of the wave
function corrections of $A_0,A_i$ and the gauge coupling $g_3^2$.
We proceed as explained in~\cite{ad}, changing $p_f\to p_f-i\mu$
in the fermion propagators, and including the 2-loop
contributions into the effective 
potential from~\cite{kaps,ka1}. Let
\ba
{\cal D}(\omega)  & = &  (4\pi)^2 \Tint{p_f}
\biggl[
\frac{1}{((p_f + i\pi T \omega)^2 + \vec{p}^2)^2}-
\frac{1}{(p_f^2 + \vec{p}^2)^2}
\biggr]  \nn \\
& = &
\int_{-\infty}^\infty \! \frac{dp}{p}\, \biggl(
\frac{1-e^p}{1+e^p}-\frac{1}{e^{p+\pi\omega}+1}+\frac{1}{e^{-p+\pi\omega}+1}
\biggr). \la{Dx}
\ea
At small $\omega$, ${\cal D}(\omega)\approx -(7\zeta(3)/2) \omega^2$. We
then obtain
\ba
g_3^2 \!\!\!\! & = & \left. g_3^2 \right|_{\mu=0}
\biggl(
1 - \frac{N_f}{6+N_c-N_f}
{\cal D}\Bigl(\frac{\mu}{\pi T}\Bigr)\, x + {\cal O}(x^2)
\biggr),  \\
x \!\!\!\! & = & \left. x \right|_{\mu=0}
\biggl(
1 - \frac{N_f}{6+N_c-N_f}
{\cal D}\Bigl(\frac{\mu}{\pi T}\Bigr)\, x + {\cal O}(x^2)
\biggr), \\
y_{N_c=2} \!\!\!\! &=&{(8-N_f)(4+N_f)\over144\pi^2x}
\biggl(1 +  \frac{3 N_f}{4+N_f}
\Bigl(\frac{\mu}{\pi T} \Bigr)^2 \biggr) \la{y_dr2} \\
& + &\frac{192-2N_f-7N_f^2-2N_f^3}{96\pi^2(8-N_f)} \biggl(
1+
\frac{3N_f(11+2 N_f)(4-N_f)}{192-2N_f-7N_f^2-2N_f^3}
\Bigl(\frac{\mu}{\pi T} \Bigr)^2 \biggr)
+{\cal O}(x),\nn \\
y_{N_c=3}\!\!\!\! &=& \!\!\!\! {(9-N_f)(6+N_f)\over144\pi^2x}
\biggl(1+ \frac{3 N_f}{6+N_f} \Bigl(\frac{\mu}{\pi T} \Bigr)^2 \biggr)
\la{y_dr3} \\
& + & \!\!\!\!
{486-33N_f-11N_f^2-2N_f^3\over96\pi^2(9-N_f)}
\biggl(1 +
\frac{3 N_f(7+N_f)(9-2N_f)}{486-33N_f-11N_f^2-2N_f^3}
\Bigl(\frac{\mu}{\pi T} \Bigr)^2 \biggr)
+{\cal O}(x), \nonumber \\
z_{N_c=3}\!\!\!\! & = & \frac{\mu}{T}\frac{N_f}{3\pi^2}
\biggl(
1 + \fr32 \frac{7+ N_f}{9-N_f}\, x + {\cal O}(x^2)
\biggr).
\ea
In the case we actually study, $N_c=3,N_f=2$, these reduce to
\ba
g_3^2 & = & \left. g_3^2 \right|_{\mu=0}
\biggl(
1 - \frac{2}{7}
{\cal D}\Bigl(\frac{\mu}{\pi T}\Bigr)\, x 
\biggr), \la{mu1} \\
x & = & \left. x \right|_{\mu=0}
\biggl(
1 - \frac{2}{7}
{\cal D}\Bigl(\frac{\mu}{\pi T}\Bigr)\, x 
\biggr), \la{xsimpl} \\
y & = & \biggl( \frac{7}{18\pi^2 x} + \frac{15}{28\pi^2} \biggr)
\biggl(1+ \frac{3}{4} \Bigl(\frac{\mu}{\pi T} \Bigr)^2 \biggr), \\
z & = & \frac{\mu}{\pi T}\frac{2}{3\pi}
\biggl(
1 + \frac{27}{14} x 
\biggr), \la{mu4}
\ea
where $g_3^2$, $\left. x \right|_{\mu=0}$
are from \eq\nr{nf2}, and corrections 
are of relative order ${\cal O}(x^2)$.
Let us recall that $x\sim 0.1$ at $T\sim 2T_c$.

Due to the small value of $x$, 
we will in the analysis which follows for simplicity ignore 
even the ${\cal O}(x)$ corrections in $g_3^2,x,z$ 
in \eqs\nr{mu1}, \nr{xsimpl}, \nr{mu4},  and use the 
leading order expressions displayed already in \eq\nr{lo-formulas}.
All of the sub-dominant
effects neglected work in the same direction 
and would strengthen the changes we see in the mass 
spectrum. For $z$ this is obvious because, 
according to \eq\nr{mu4}, for a given $\mu/T$ 
the physical value would be larger than what we have used.
For $x$,
\eq\nr{xsimpl} and the negative value of ${\cal D}$ imply that
the fact that we have kept $x$ unchanged means that we have 
effectively moved slightly up in $T/\lambdamsbar$. 
In $g_3^2$ a major part of this effect cancels since
the correction is the same in \eqs\nr{mu1}, \nr{xsimpl}.  
Therefore, $M/T$ remains essentially the same for given $M/g_3^2$, 
even though the temperature is slightly higher. 

\subsection{A complex action and reweighting}
\la{coma}

The generic problem of SU(3) simulations at non-vanishing 
chemical potential now appears as follows. 
For $\mu\neq 0$, there is the imaginary
term $i z S_z$ in the action, and the conventional importance
sampling will not work\footnote{It can be seen directly
in 4d that the sign problem is strongly correlated with the 
imaginary part of the temporal Polyakov loop~\cite{ippl}.
This corresponds precisely to $z S_z$ in the 3d language.}. 
A way to circumvent this is to do
the importance sampling with the action at $z=0$, 
while $\exp(-i z S_z)$ is included in the operators. 
We will refer to this procedure as ``reweighting''.
Reweighting is also applied in the ``Glasgow method'' suggested 
for $\mu\neq 0$ simulations in 4d (for a review, see~\cite{gl});
however, in our approach we do not need 
to carry out any expansion related to $z$.
Denoting by brackets the expectation value with respect to
the action in \eq (\ref{lattice_action}),
operators even $O_{+}$ and odd $O_{-}$ in $R$ 
therefore appear as
\bea
\langle O_{+} \exp(-i z S_z) \rangle =
\langle O_{+} \cos(z S_z) \rangle, \quad
\langle O_{-} \exp(-i z S_z) \rangle = -i
\langle O_{-} \sin(z S_z) \rangle. \la{meas}
\eea
Since \eq\nr{meas} involves trigonometric functions, 
the problem manifests itself if configurations with $S_z \gg 1$
occur frequently. In this case there are cancellations in the functional
integral, and the Monte Carlo method will soon lose its accuracy.

Configurations with $S_z \gg 1$
need not always be typical, however. Indeed, let us
consider the distribution of $S_z$ in \eq\nr{sz}
for $z=0$. The centre of the distribution is at zero.
The fact that there is Debye screening, $y > 0$, guarantees
that the distribution is approximately Gaussian. Its width
is easily computed, and is 
given by the 2-loop scalar sunset graph on the lattice,
which has been evaluated in~\cite{framework,contlatt}: 
\beq
\Delta (z S_z)
\approx \frac{z}{4\pi}\Bigl(\frac{6 L}{\beta}\Bigr)^{3/2}
\biggl[5\biggl(
\ln \frac{\beta}{3y^{1/2}}+0.58849
\biggr)\biggr]^{1/2}. \la{osccond}
\eeq
We note that the width grows
with the lattice size $L$ (or, more precisely,
with the physical lattice size $L/\beta$)\footnote{For a fixed
physical lattice size, the width grows with $\beta$, but only
logarithmically.}. 
Nevertheless, if $z$ is small enough,
the infinite volume and continuum scaling limits of the 
correlation functions may be extracted before \eq\nr{osccond}
grows to values $\gg 1$. In these cases we can in practice carry
out all the measurements before the oscillations set in.
The feasibility of this procedure can be checked by
monitoring the numerical distribution of $z S_z$,
the argument of the reweighting factors. Accurate calculations should
be possible whenever the bulk of the distribution is contained 
in the interval $(-\pi,\pi)$.
Once the distribution gets significantly broader,
the breakdown should show up in the correlation functions 
simply as a noisy signal.

Even assuming that the oscillations are under control, reweighting 
according to \eq\nr{meas} of course requires some care. The importance
sampling of the Monte Carlo is dominated by the minimum of the action
at $z=0$, which is unmodified by including $S_z$ in the operators.
If the $z=0$ minimum is widely separated from the true $z\neq0$ minimum
in configuration space, there is a danger
of biasing the Monte Carlo towards the wrong configurations.
In order to check this, we have tested the reweighting method
by considering $z$ imaginary, $z\to -i \zeta$.
In this case the action is real and the scalar update can be 
modified by a Metropolis step to include $S_z$, and hence update
with the exact action.
The results can then be compared with
those obtained from \eq\nr{meas}, now with
\beq
\cos(z S_z)\to\cosh(\zeta S_z), \quad
-i \sin(z S_z)\to-\sinh(\zeta S_z).
\eeq
Simultaneously, we change the sign of $\mu^2$ in $y$
in \eq\nr{lo-formulas}.

Finally, 
as a small technical detail let us note that 
in practice we implement $S_z$ in \eq\nr{sz}
on the lattice by including
an ${\cal O}(a)$ improved~\cite{moore_a} version of 
$\tr A_0^3$, as in \eq(2.10) of~\cite{su3adj}.

\subsection{Determining masses with a complex action}

Assuming that our reweighting procedure works, we then 
have to extract masses in the changed symmetry situation.
After the inclusion of $\mu\neq 0$, the action of the effective theory
is no longer real, and no longer invariant in $R, C$. It is
however still invariant
under $RC$ and $P$. In the scalar channel this means that the
operators $0^{++}_{+}$ and $0^{+-}_{-}$ in \eq\nr{contop}
can couple to each
other, and similarly $0^{--}_{+}$ and $0^{-+}_{-}$.
These two channels
are still distinguished, however, by the parity~$P$.

We measure the correlation matrix between the two
channels which used to be decoupled for $\mu=0$. {}From \eq\nr{meas}
we know that the result is of the block form
\beq
C(t) = \left(
\begin{array}[c]{cc}
A(t) & -i z D(t) \\
-i z D(t) & B(t)
\end{array}
\right), \la{imC}
\eeq
where $A(t)$, $B(t)$ are symmetric matrices (possibly of
different sizes), representing the correlations within the
two channels as in \eq\nr{cij}. 
We have factored out $z$ in the off-diagonal
blocks, to make it clear that the result is odd in $z$ and
vanishes for $z\to 0$. 
In order to facilitate the solution of the eigenvalue problem 
in \eq (\ref{evp}), we may note that we
can equivalently consider the eigenvalue problem for the 
real correlation matrix
\beq
\tilde C(t) = \left(
\begin{array}[c]{cc}
A(t) & -z D(t) \\
z D(t) &  B(t)
\end{array}
\right), \la{reC}
\eeq
obtained from $C(t)$ by a similarity transformation $Z=\mbox{diag}\,(1,i)$.
Finally, for imaginary chemical potentials $z=-i\zeta$ we have
a correlation matrix of the usual symmetric form
as in \eq\nr{cij},
\beq
C_\zeta(t) = \left(
\begin{array}[c]{cc}
A_\zeta(t) & -\zeta D_\zeta(t) \\
-\zeta D_\zeta(t) & B_\zeta(t)
\end{array}
\right). \la{reD}
\eeq

In addition, let us note 
that physical observables such as masses (i.e., inverse correlation lengths)
must be even in $z$, since a change of its sign can be compensated for
by the field redefinition $A_0\to -A_0$
in the action in \eq\nr{lo-formulas}. Since the mass spectrum
is well defined and there are no massless modes at $z=0$, we 
moreover expect that the system is analytic in $z$ for small $|z|$,
the expressions of the masses starting with $z^2$. In particular,
masses must remain real for small enough $z$
(even though for the eigenvalues of a fully general matrix
of the form in \eqs\nr{imC}, \nr{reC} this is not always the case).

We can use these observations to discuss the form of the mass eigenstates.
For the cases in \eqs\nr{imC}, \nr{reC}, we do not in general have
orthonormal eigenvectors in the usual sense. 
In order for the procedure
to make physical sense, we however expect (and observe) that the 
eigenvectors are independent of $t$ within statistical accuracy,
with eigenvalues of the form $\sim \exp(- a M t)$, 
as in \eq\nr{evp}.

Consider now first \eq\nr{reD}. Since the eigenvalues
are even in $\zeta$, the eigenvectors should be of the forms
\beq
\vec{v}^{(1)} = \left(
\begin{array}[c]{c}
\vec{v}_A \\
\zeta \vec{v}_B
\end{array}
\right), \quad
\vec{v}_A^2 +
\zeta^2 \vec{v}_B^2 = 1; \quad
\vec{v}^{(2)} = \left(
\begin{array}[c]{c}
- \zeta \vec{v}_A \\
\vec{v}_B
\end{array}
\right), \quad
\zeta^2 \vec{v}_A^2 +
\vec{v}_B^2 = 1, \la{nor_reD}
\eeq
where normalisation conditions have also been shown. 
With analytic continuation, 
we can then directly apply this to the
case of \eq\nr{imC},
\beq
\vec{v}^{(1)} = \left(
\begin{array}[c]{c}
\vec{v}_A \\
i z \vec{v}_B
\end{array}
\right), \quad
\vec{v}_A^2 -
z^2 \vec{v}_B^2 = 1; \quad
\vec{v}^{(2)} = \left(
\begin{array}[c]{c}
- i z \vec{v}_A \\
\vec{v}_B
\end{array}
\right), \quad
- z^2 \vec{v}_A^2 +
\vec{v}_B^2 = 1,
\eeq
as well as to the case of \eq\nr{reC}, 
using the similarity transformation:
\beq
\tilde{\vec{v}}^{(1)} = \left(
\begin{array}[c]{c}
\vec{v}_A \\
-z \vec{v}_B
\end{array}
\right), \quad
\vec{v}_A^2 - z^2
\vec{v}_B^2 = 1; \quad
\tilde{\vec{v}}^{(2)} = \left(
\begin{array}[c]{c}
-z\vec{v}_A \\
\vec{v}_B
\end{array}
\right), \quad
- z^2 \vec{v}_A^2 +
\vec{v}_B^2 = 1. \la{nor_reC}
\eeq
We observe that the ``metric'' has changed, but otherwise
the situation is quite analogous to the usual one
in \eq\nr{nor_reD}.

Fortunately, in practice even less needs to be changed
in the usual numerical analysis based on a matrix of the form
in \eqs\nr{cij}, \nr{reD}, 
if we use \eq\nr{reC}. There are algorithms finding the eigenvalues
and eigenvectors for a general non-symmetric real matrix. The
only change is in the normalisation according to \eq\nr{nor_reC}, 
which amounts to fixing an overall coefficient
for each eigenvector.  However, 
as we only consider the
correlations between the eigenvectors, 
cf.\ \eq\nr{Mt}, and do not need
to verify orthogonality explicitly, we can
equally well employ a ``wrong'' scalar product 
and normalisation based on the old type of metric 
in \eq\nr{nor_reD} (with $\zeta \to z$). Then everything goes
precisely as before.

Finally, let us illustrate the general expectations
for the mass pattern with a simple $2\times 2$ matrix
as an analogue of \eq\nr{imC}. Consider two
real scalar fields $\phi,\chi$, with a mass term
\beq
V = \fr12 m^2 \phi^2 + \fr12 M^2 \chi^2 + i\epsilon \phi\chi. \la{tm}
\eeq
For $M\gg m$ and $\epsilon < (M^2-m^2)/2$, 
the mass eigenvalues are real, $m^2+(\epsilon/M)^2$,
$M^2-(\epsilon/M)^2$. 
Thus the lightest mass goes up, which is 
what we expect for a real chemical potential. In fact 
there is another effect contributing in the same
direction, since according to \eq\nr{lo-formulas}
the parameter $y$ grows with $\mu/T$.
In the case of a complex $\epsilon=-i\eta$, on the
contrary, the lightest mass becomes even lighter,
the heaviest mass even heavier: $m^2-(\eta/M)^2$,
$M^2+(\eta/M)^2$. Again, the change of $y$
contributes in the same direction for the lightest mass.

\section{Numerical results for $\mu\neq 0$}

We now proceed to present the results of our first numerical explorations
of $\mu\neq 0$.
We restrict our attention to the case $T=2T_c$, $N_f=2$, corresponding
to $x=0.0919$. As we have
explicitly checked for good scaling behaviour at 
$\mu=0$, we work exclusively with $\beta=21$ in this section.
The parameters and lattices considered are summarised in Table 
\ref{tab_params_mu}, and the detailed results
are displayed in Tables~\ref{tab_0_immu}, \ref{tab_0_remu}
in the Appendix.
\begin{table}[tb]
\begin{center}
\begin{tabular}{|r@{.}l|r@{.}l|r@{.}l|l|}
\hline
\hline
\multicolumn{2}{|c|}{$\mu/T$} &
\multicolumn{2}{c|}{$y$} &
\multicolumn{2}{c|}{$z$} & $L^2 \cdot T$ \\
\hline
 0&5 \ii &  0&47382 & 0&0338 \ii & $20^3$ \\
 1&0 \ii &  0&44630 & 0&0675 \ii & $20^3$ \\
 1&5 \ii &  0&40042 & 0&1013 \ii & $30^3$ \\
 0&5    &  0&49218 & 0&0338 & $30^3$ \\
 1&0    &  0&51970 & 0&0675 & $30^3$ \\
 1&5    &  0&56558 & 0&1013 & $30^3$ \\
 2&0    &  0&62981 & 0&1351 & $30^3, 40^3$ \\
 4&0    &  1&07026 & 0&2702 & $10^3,14^3,18^3,30^3$ \\
\hline 
\hline
\end{tabular}
\caption{ \label{tab_params_mu}
{\em The lattice parameters and sizes used for calculations with 
$\mu\neq 0$. All are for $T=2\lambdamsbar, N_f=2$, $x=0.0919$, $\beta=21$.
The parameter $z$ is from \eq\nr{lo-formulas}.}}
\end{center}
\end{table}

\subsection{Consistency checks and the general pattern}

\begin{figure}[tb]%

\vspace*{-3cm}

\begin{center}
\leavevmode
\epsfxsize=200pt
\epsfbox[20 30 620 730]{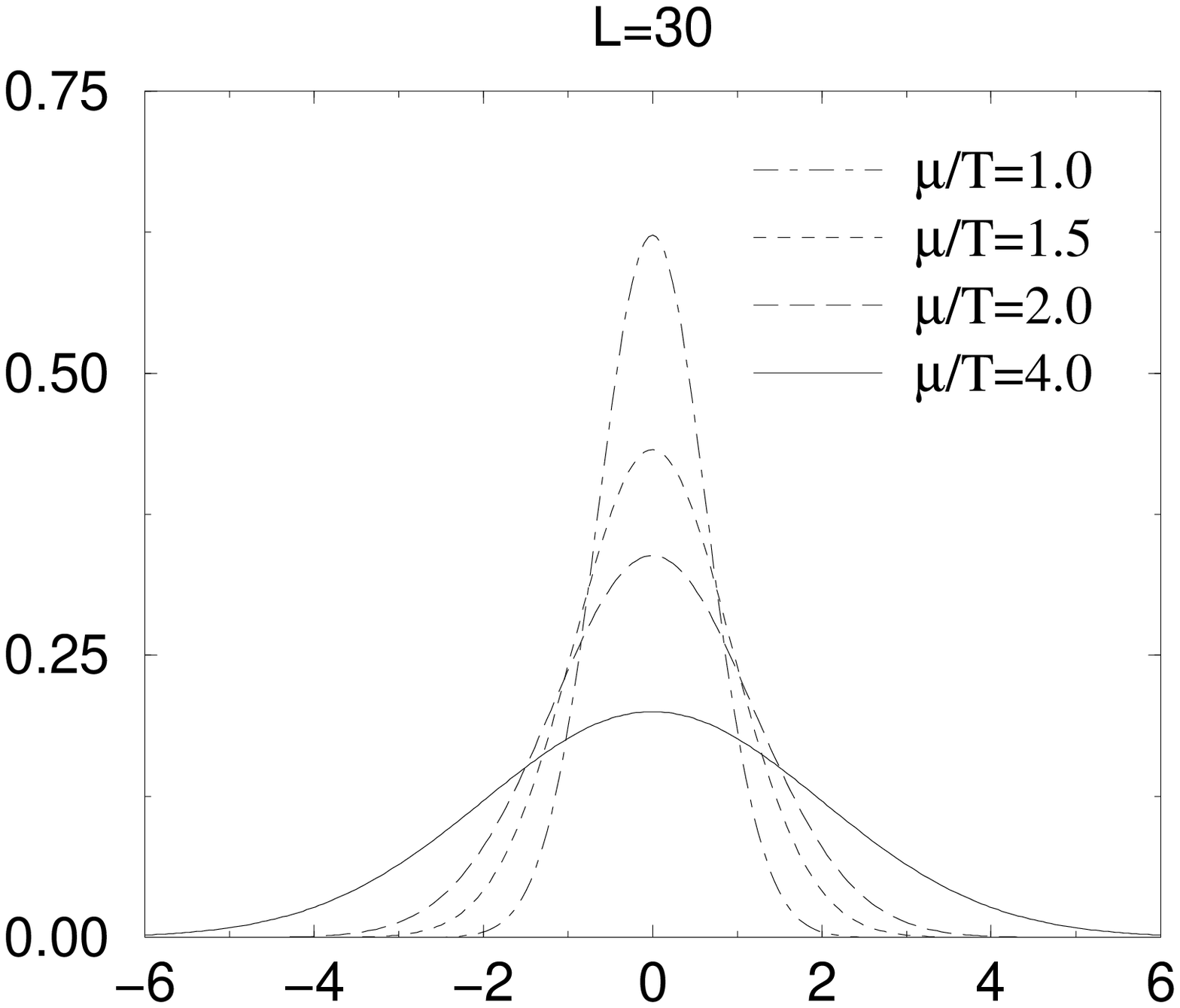}
\leavevmode
\epsfxsize=200pt
\epsfbox[20 30 620 730]{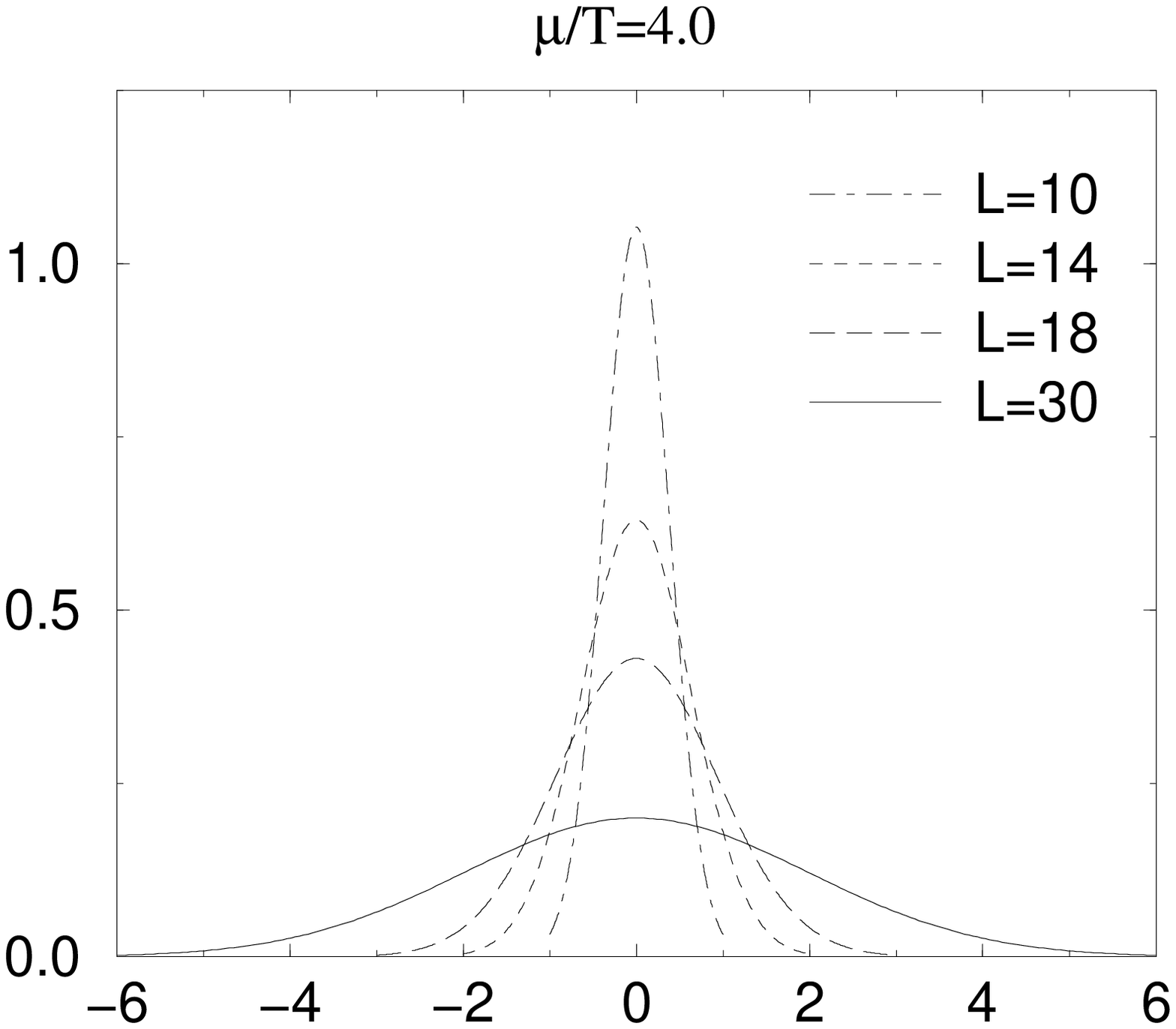}
\leavevmode
\end{center}

\vspace{-1.0cm}

\caption[]{\label{hist}{\it
Distribution of $zS_z[A_0]$ for the reweighting procedure
(\eq\nr{meas}).}}
\end{figure}

We begin by assessing the distribution of the argument of the
reweighting factors in \eq\nr{meas}. We find that any typical 
histogram can be easily fitted to a
Gaussian, from which we determine its width.  The
plots in \fig \ref{hist} show the 
corresponding distributions as a function of $\mu/T$
and the lattice volume, with behaviour as expected from \eq
(\ref{osccond}).  The volume chosen for the $\mu/T$ series is large
enough to be free from finite size effects in the $\mu=0$ simulations.
As the histograms show, for these volumes the distribution of the
argument of the reweighting factor is entirely contained within one
period for $\mu/T\leq 2.0$. On the other hand, 
at $\mu/T=4.0$ the tails of the
distribution are significantly spreading.  Nevertheless, the bulk of the
distribution is still within the region without sign flips.  For the
same number of measurements at $\mu/T=2.0,4.0$, however, the errors on
the correlation functions are indeed by a factor of three larger in the
latter case, confirming the onset of the sign problem.  Going to still
larger $\mu/T$ will thus lead to increasingly noisy signals, until a
mass determination becomes impossible.

\begin{figure}[tb]%

\vspace{-3.0cm}

\begin{center}
\leavevmode
\epsfxsize=200pt
\epsfbox[20 30 620 730]{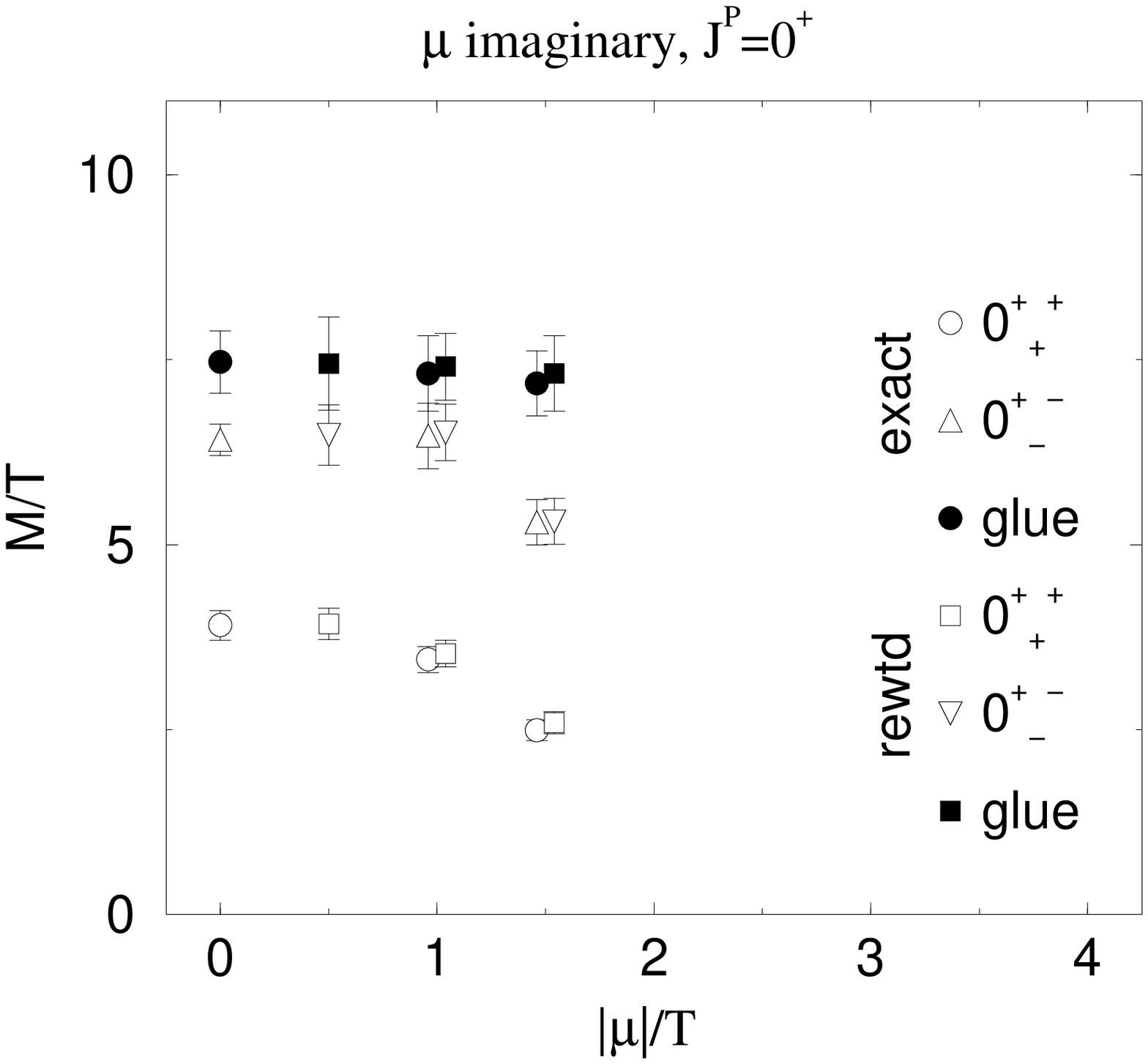}
\leavevmode
\epsfxsize=200pt
\epsfbox[20 30 620 730]{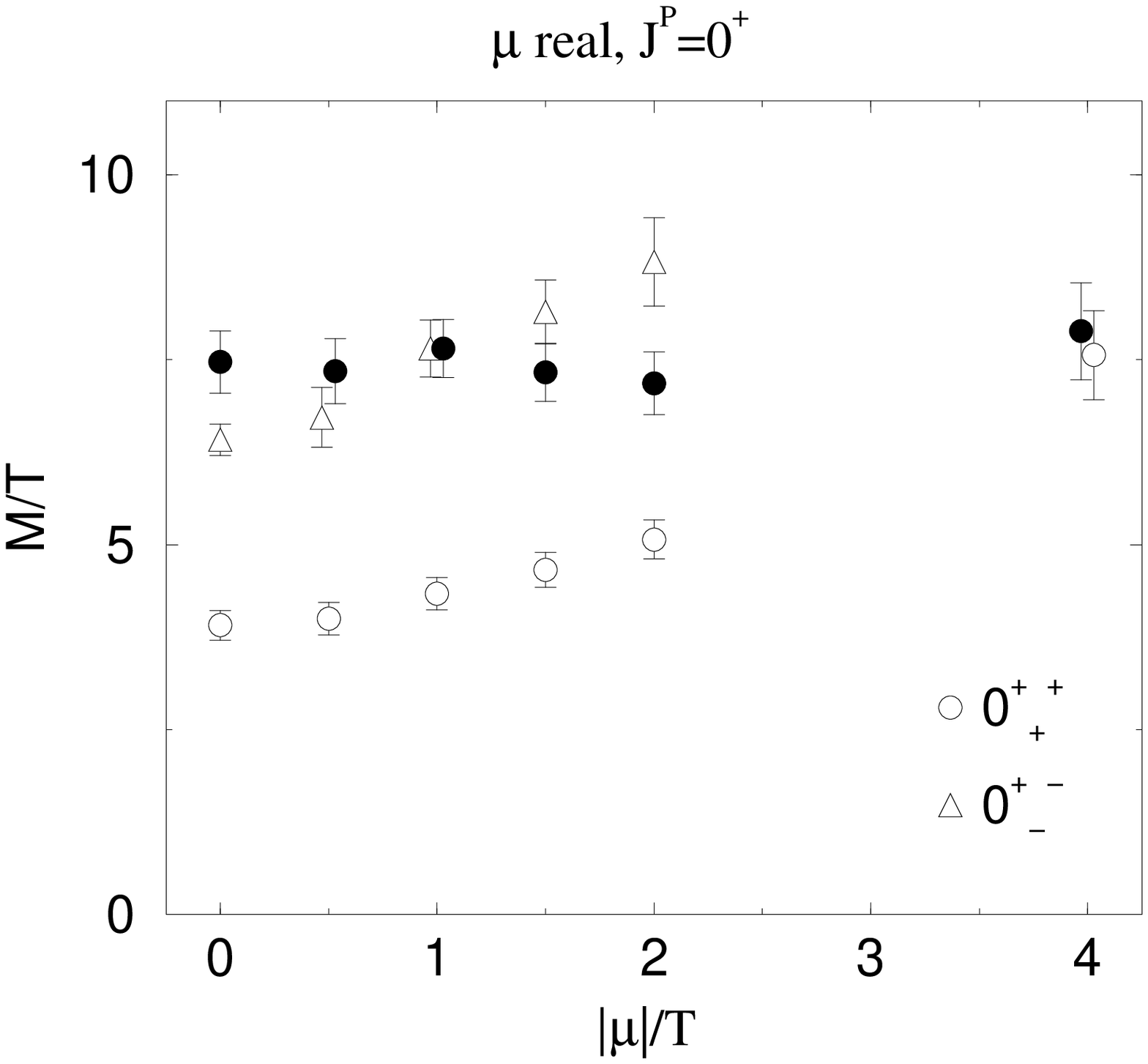}
\leavevmode
\end{center}

\vspace{-4.0cm}

\begin{center}
\leavevmode
\epsfxsize=200pt
\epsfbox[20 30 620 730]{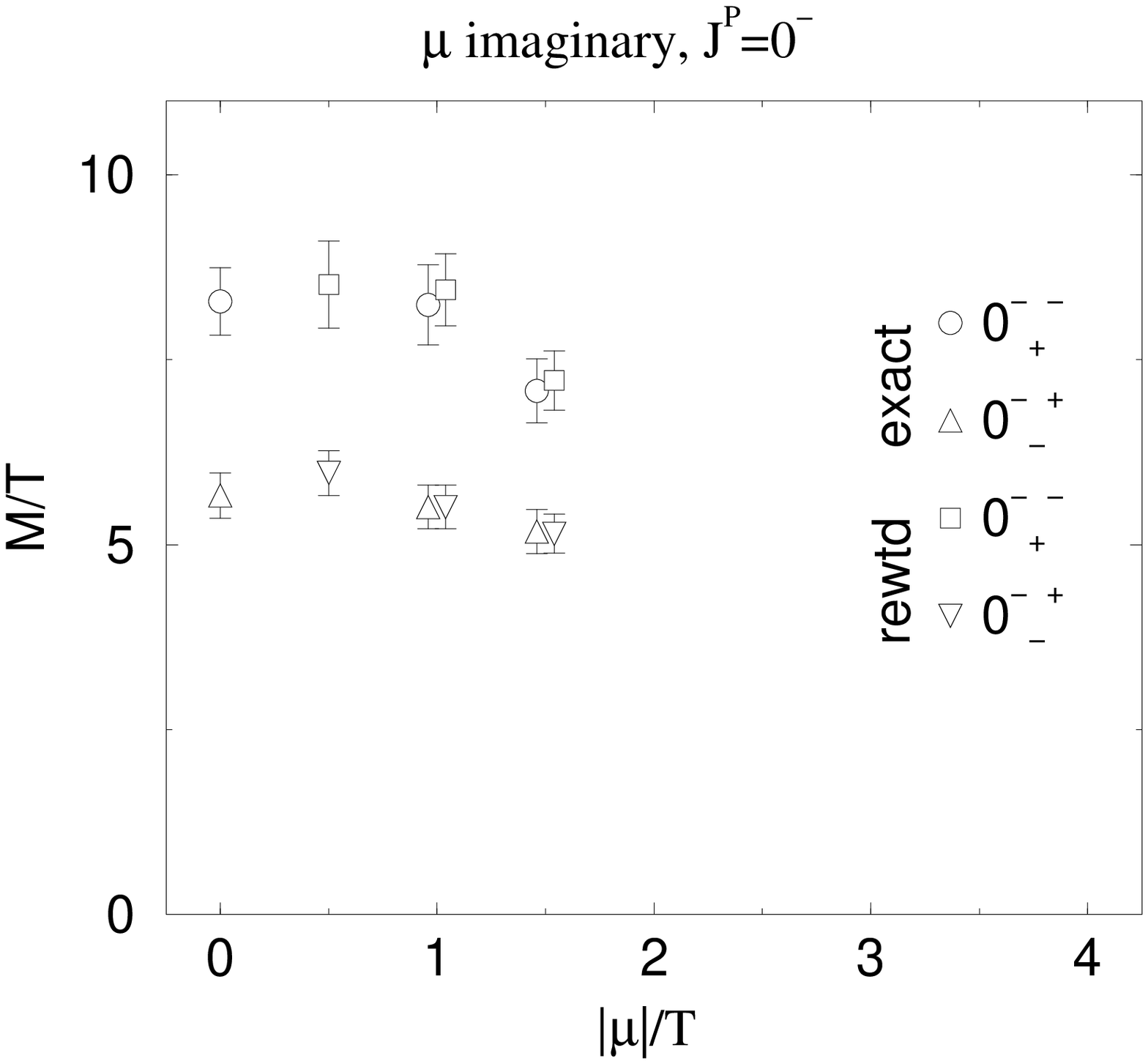}
\leavevmode
\epsfxsize=200pt
\epsfbox[20 30 620 730]{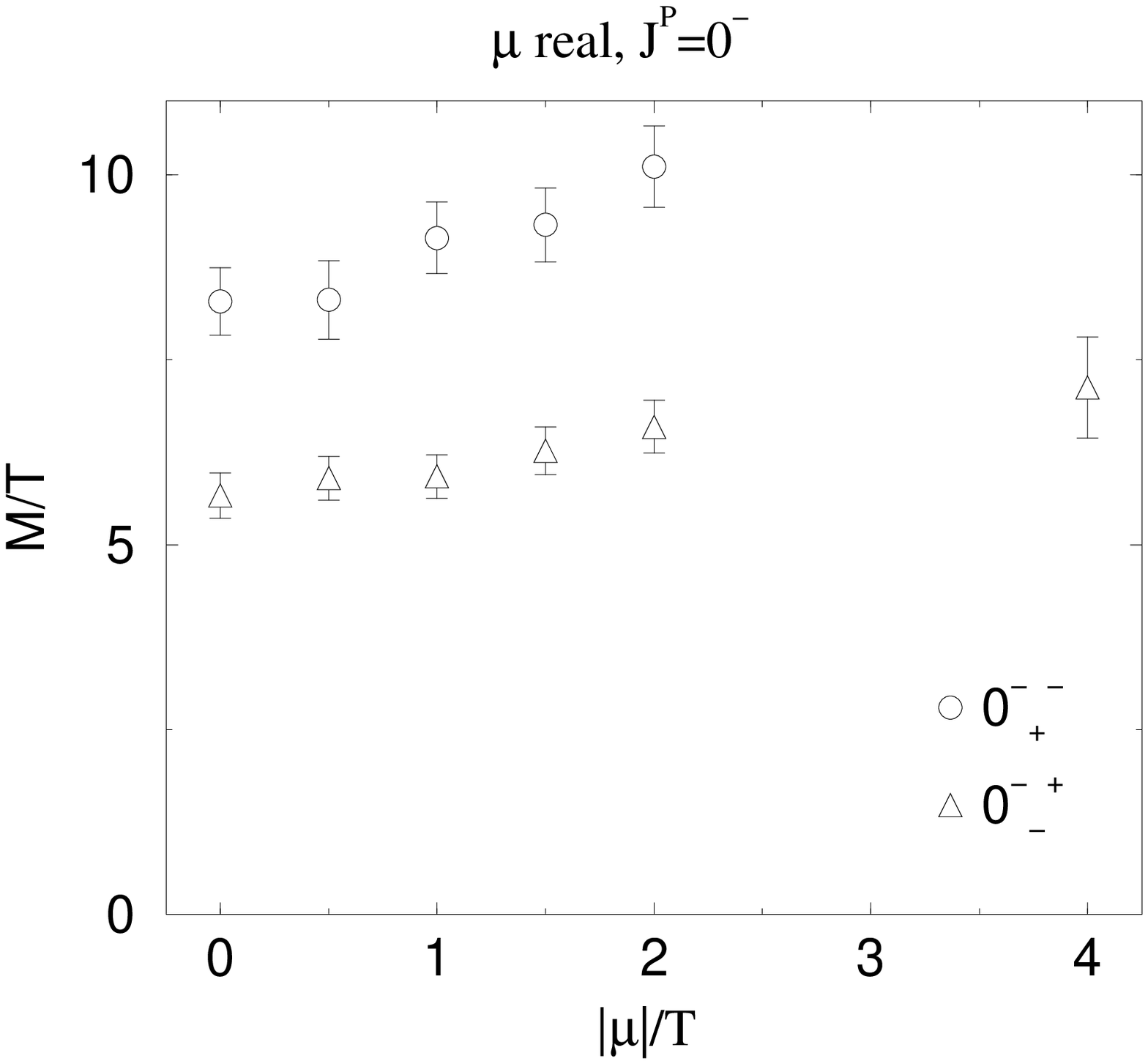}
\leavevmode
\end{center}

\vspace{-1.0cm}

\caption[]{\label{specmu}{\it
    The dependence of the lowest thermal masses on $\mu/T$ for
    imaginary (left) and real (right) chemical potential. For clarity
    of presentation, we show the lightest state only for each quantum
    number assignment. The quantum number assignments refer to the 
    eigenstates at $\mu/T=0$. Filled symbols correspond to glueballs.
    ``Exact'' and ``reweighted'' simulations (left) are explained 
    in Sec.~\ref{coma}.}}
\end{figure}

In the next step, we test the reliability of the reweighting procedure
by comparing with the correct importance sampling in the case of an
imaginary chemical potential. 
This comparison can only be carried out for a limited range of
$\mu/T$. In the full action, the imaginary chemical potential term
weakens the metastability~\cite{ad} of the symmetric phase, and tunnelling
during typical simulations to the unphysical phase becomes
possible. This can be countered by increasing the lattice volume,
but in so doing we increase the width of the distribution of
$zS_z[A_0]$. We find however that
there exists a window of opportunity between these two limits where we
can usefully compare the two algorithms.
In the left panels of
\fig \ref{specmu}, the ground
states of the $0^+$ and $0^-$ channels are shown as a function of
imaginary $\mu/T$. First, we note the correct qualitative behaviour of 
a mass decrease with $|\mu/T|$, as described
after \eq\nr{tm}. Second, we observe complete
agreement within the small statistical errors 
between the reweighted observables and those measured
with the full action.

With these tests passed, we can thus display the same states
for the case of real chemical potential in
the right panels of \fig \ref{specmu}. Again
in agreement with the qualitative behaviour expected from 
the discussion after \eq\nr{tm}, the lowest mass
values are growing with increasing $\mu/T$, 
since the constituent mass parameter $y^{1/2}$
gets larger, see \eq\nr{lo-formulas}. Thus the corresponding
correlation lengths decrease. We note again, however, that 
purely gluonic states 
(or, in the 4d language, purely magnetic states $\sim \tr F_{12}^2$)
are rather insensitive to $\mu/T$, in analogy with their behaviour
under variations of $y$ discussed earlier.  This means that at large
enough $\mu/T > 4.0$, they may become the lightest states
in the system, as suggested by the top right panel of \fig\ref{specmu}.

At $\mu/T=2.0$ we have performed a finite volume check, comparing
lattices of $L=30$ and $40$ at $\beta=21$. We find the low lying
masses to be consistent within statistical errors, 
see Table~\ref{tab_0_remu}. This is
encouraging, implying that the reweighting does not magnify finite
volume effects unduly, and that it is indeed possible to extrapolate
screening masses to the infinite volume in a finite chemical potential
context, before the onset of oscillations.

\subsection{The Debye screening length at finite baryon density}

Let us now consider what happens to 
the Debye screening length as defined in \se\ref{debye}.
This definition is based 
on the $R$-symmetry of our effective theory~\cite{ay}. 
For $\mu\neq 0$ the extra term in the action 
spoils the symmetry, resulting in a merging of the various $RC$-channels,
while only $P$ remains as a good quantum number. The question then arises,
how can the Debye mass be defined in such a situation~\cite{ay}?

We now note the following.  For $\mu=0$, the lightest $R=-1$ state was
$0_-^{-+} \sim \tr A_0 F_{12}$. This is, however, also the lightest
state in the channel $P=-1$. Thus, even after $\mu\neq 0$, the same
operator continues to determine the long distance decay of a number of
$R=-1$ operators.  The values can be seen in \fig\ref{specmu}, bottom
right.  If we {\em define} this largest $P=-1$ correlation length to
correspond to Debye screening, then $\mu\neq 0$ does {\em not} change
the situation in an essential way, and the corresponding correlation
length decreases with~$\mu$. An example of a 4d gauge invariant
operator with such a behaviour is $\tr F_{03} F_{12}$~\cite{ay}.

On the other hand, the  $0_-^{+-}$ operator $\sim \tr A_0^3$ 
does couple to $0_+^{++}\sim \tr A_0^2$ after the inclusion of 
$\mu\neq 0$. Therefore, its correlation length strictly 
speaking {\em increases} by a factor of almost two as soon as 
$\mu\neq 0$, as can be seen from the top right panel
in \fig\ref{specmu}. There we have labelled the states with
the quantum numbers of the $\mu/T=0$ limit; however, for
$\mu/T> 0$, any operator which originally had just the quantum
numbers $0_-^{+-}$ such as $\tr A_0^3$, 
couples now also to $0_+^{++}$. The eigenstate
shown in \fig\ref{specmu} which does have a 
larger mass, is a particular linear combination of the 
two types of original
states. Nevertheless, the overlap between the two operators 
$\sim \tr A_0^2$, $\sim \tr A_0^3$ is very small for small $\mu/T$, 
so that even for $\sim \tr A_0^3$ there is an intermediate 
distance range where correlations should decay as
shown with $0_-^{+-}$ in~\fig\ref{specmu}, 
as discussed in~\cite{ay}. 
Thus no abrupt change is observed.
An example of a gauge invariant 4d operator 
with such a behaviour is the imaginary 
part of the 4d temporal Polyakov loop. 

\section{Conclusions}

In this paper, we have carried out mass
measurements in the different quantum number channels 
within the 3d SU(3)+adjoint Higgs theory. 
We have stayed on the dimensional reduction curve
in the symmetry restored phase of the 3d theory, 
and the results are thus 
expected to correspond to spatial correlation lengths 
in the deconfined quark--gluon plasma phase of QCD. 
This interpretation appears to be relatively
accurate at least down to $T\sim 2T_c$.  We first considered
a vanishing chemical potential $\mu=0$, 
and then also extended 
the measurements to a finite $\mu/T\lsim 4.0$.

For $\mu=0, N_f=0$, we believe that the asymptotic correlation lengths
related to the real and imaginary parts of the 4d temporal Polyakov 
loop, as well as to other gauge invariant bosonic operators, 
are now relatively well understood. For instance, at 
$T\sim 2T_c\approx 2\lambdamsbar$,  
the real part of the Polyakov loop (with quantum numbers $0_+^{++}$
according to our conventions discussed in \se\ref{qn}) 
decays exponentially at $\sim (3 T)^{-1}$, while the 
imaginary part ($0_-^{+-}$) decays at $\sim(5 T)^{-1}$, 
as can be observed from \fig\ref{spec_nf0}. The Debye
screening length according to the definition of~\cite{ay}
turns out to be determined by operators of the type 
$0_-^{-+}\sim \tr F_{03} F_{12}$, and 
is slightly longer than the screening length of the
imaginary part of the Polyakov loop, $\sim(4.6 T)^{-1}$. 

We have also explicitly studied the effects of $N_f$ dynamical fermions
on the longest correlation lengths. 
For fixed $T/\lambdamsbar$, the correlation lengths decrease as $N_f$
is increased.  For instance,  in the phenomenologically
interesting case of $N_f = 2,3$ at $T\sim 2\lambdamsbar$, 
the real part of the 
Polyakov loop decays exponentially at $\sim (4T)^{-1}$, 
while the imaginary part at $\sim (6T)^{-1}$.
As a theoretical point, it is interesting to note
that in units of $g_3^2$,
the screening lengths scale quite well with $N_c$. 

We have then extended the measurements to $\mu\neq 0$. We have 
demonstrated that the use of dimensional reduction allows one to 
carry out simulations corresponding to 
phenomenologically interesting values of 
quark masses, temperatures, and $\mu/T$ (although 
not at the point of the phase transition).  
Simulations are possible because the only term in the 
super-renormalizable 3d action 
suffering from the sign problem, 
$\sim i \tr A_0^3$, comes with a very small
numerical coefficient. 

In general, we observe that as $\mu/T$ is switched on at a fixed
temperature, the screening lengths decrease. This can be interpreted
so that the lightest excitations, involving $A_0$, are ``less  critical'' 
and correspondingly heavier. Thus, staying at a fixed temperature
but increasing $\mu$, we are apparently moving further away from 
the critical line, which hence has to bend down. This is in complete 
accordance with what we qualitatively know about the phase diagram 
in the $(\mu,T)$-plane.

However, there is one state whose mass does not increase, the
3d glueball $\sim\tr F_{12}^2$. Thus at large enough $\mu/T$, 
similarly to the case of large enough $T/\lambdamsbar$ at $\mu=0$, 
it becomes the lightest excitation in the system. We find
that the crossover corresponds roughly to $y\sim 1.1$
in terms of the dimensionless parameter defined 
in \eqs\nr{xy}, \nr{lo-formulas}. For $N_f=2$, this 
corresponds roughly to $T/\lambdamsbar \sim 10^2$ at $\mu/T\sim 0.0$, 
and $\mu/T\sim 4.0$ at $T/\lambdamsbar \sim 2$.

Finally we have addressed the behaviour of the imaginary 
part of the 4d temporal Polyakov loop. As the chemical 
potential is switched on, its asymptotic correlation length
suddenly increases contrary to the other observables discussed, 
and agrees for all $\mu\neq 0$
with that of the real part of the 4d temporal Polyakov loop.

The task remains to establish a direct connection between 
the screening masses discussed here and observables relevant 
for the phenomenology of heavy ion collision experiments. 

\section*{Acknowledgements}

We acknowledge useful discussions with R.D. Pisarski and K. Rajagopal. 
This work was partly supported by the TMR network {\em Finite
  Temperature Phase Transitions in Particle Physics}, EU contract no.\ 
FMRX-CT97-0122. The work of A.H. was supported in part by UK PPARC
grant PPA/G/0/1998/00621.

\section*{Note added}

After the submission of our paper, a letter appeared by 
Gavai and Gupta~\cite{gg}, who measure the spatial meson (pion) 
correlation length at $T\sim (1.5...2.9) T_c$ for $N_f=4$ light 
dynamical fermions. They observe that at $T\sim 2T_c$ the meson 
``mass'' measured from $\bar\psi\psi$ has become lighter
than the lightest bosonic mass (represented by
$\sim \tr A_0^2$ in the 3d language), 
so that the bosonic gauge degrees of freedom are no longer the lightest
dynamical degrees of freedom, and thus the standard dimensionally 
reduced effective theory should not be reliable any more. This observation 
is in accordance with our comments in Sec.~\ref{magsec}. However, Gavai 
and Gupta make the further suggestion that even in the temperature
range $T\sim (2...10) T_c$ where the meson mass is no 
longer the lightest one, but is still below its asymptotic
value ($\sim 2\pi T$ in the case of the chiral continuum limit), 
dimensional reduction may not be reliable, 
in contrast to the situation in the purely bosonic case $N_f=0$.
Let us comment that there is no hard evidence for
the latter suggestion. In fact, it seems quite possible 
to us that the decrease from $\sim 2\pi T$
could be partly accounted for precisely by the confining bosonic 3d
gauge field dynamics related to $A_0,A_i$. Furthermore, we see
no evidence for why the bosonic correlation lengths we have 
measured here would not be satisfactorily reproduced by the 3d theory
in this temperature range. Thus we believe that 
our bosonic mass measurement for $N_f>0$ (both at $\mu/T=0$ 
and $\mu/T>0$) do reproduce at least the qualitative features of 
the physical 4d theory at all temperatures $T\gsim 2T_c$. 

As already discussed by Gavai and Gupta, 
one might also expect that 
chiral effects responsible for the behaviour
seen by them are at $N_f=4$ somewhat larger than in the
realistic case $N_f=2...3$.
For instance, $\tr A_0^2$ is lighter for smaller $N_f$ as we have
discussed in this paper, while $\bar\psi\psi$ still reaches the 
same asymptotic value, so that the level crossing invalidating
dimensional reduction should take place at a smaller temperature.

\appendix
\renewcommand{\thesection}{Appendix~~\Alph{section}}
\renewcommand{\thesubsection}{\Alph{section}.\arabic{subsection}}
\renewcommand{\theequation}{\Alph{section}.\arabic{equation}}

\section{Tables}

For completeness, 
we collect in this Appendix the numerical results for 
the correlation length measurements we have carried out.

\begin{table}[htb]
\begin{center}
\begin{tabular}{|l||r@{.}lc|r@{.}lc||r@{.}lc||r@{.}lc|}
\hline
          &
\multicolumn{9}{|c||}{$T=2T_c$} &
\multicolumn{3}{c|}{$T \sim 10^{11}T_c$} \\
\cline{2-13}
&
\multicolumn{6}{|c||}{$\beta=21$}  &
\multicolumn{3}{|c||}{$\beta=28$}  & 
\multicolumn{3}{|c|}{$\beta=28$} \\
\cline{2-13}
          &
\multicolumn{3}{|c|}{$L=38$} &
\multicolumn{3}{|c||}{$L=28$} &
\multicolumn{3}{|c||}{$L=40$} &
\multicolumn{3}{|c|}{$L=40$} \\
\hline
\hline
$0^{++}_+$  &  \multicolumn{2}{c}{$M/g_3^2$} & Ops.   &
\multicolumn{2}{c}{$M/g_3^2$} & Ops.  &
\multicolumn{2}{c}{$M/g_3^2$} & Ops.  &
\multicolumn{2}{c}{$M/g_3^2$} & Ops.  \\
\hline
$\Phi_1$ & 
1&022(15)   & $R,L$     & 
1&022(11)   & $R,L$     & 
0&994(19)   & $R,L$     &
2&55(13)   & $C$       
\\
$\Phi_2$ & 
2&44(14)   & $C$     & 
2&51(11)   & $C$     & 
2&511(66)   & $C$     &
2&86(14)   & $P$     
\\
$\Phi_3$ & 
2&92(14)   & $(R,L)$     & 
2&62(11)   & $T$     & 
2&65(8)   & $T$     &
3&75(21)   & $C$     
\\
$\Phi_4$ & 
3&70(18)   & $T$     & 
2&95(14)   & $(L)$     & 
2&95(16)   & $(R,L)$   &
4&28(28)   & $(P)$     
\\
$\Phi_5$ & 
3&79(18)   & $C$     & 
3&66(21)   & $C$     & 
3&61(32)   & $C$     &
4&89(27)   & $R,L$   
\\
$\Phi_6$ & 
4&62(28)   & $(R,L)$     & 
4&14(25)  & $(T)$ &
3&66(36)   & $(T)$    &
\multicolumn{3}{c|}{---}
\\
\hline
\hline
$2^{++}_+$  &  \multicolumn{2}{c}{$M/g_3^2$} & Ops.   &
\multicolumn{2}{c}{$M/g_3^2$} & Ops.  &
\multicolumn{2}{c}{$M/g_3^2$} & Ops.  &  
\multicolumn{2}{c}{$M/g_3^2$} & Ops.  \\
\hline
$\Phi_1$ & 
3&60(18)   & $T$     & 
2&63(7)   & $T$     & 
2&82(13)   & $T$     &
2&94(17)   & $T$     
\\
$\Phi_2$ & 
3&55(19)   & $R,L$     & 
3&44(16)   & $R,L$     & 
3&355(94)   & $R,L$     &
4&26(19)   & $C$     
\\
$\Phi_3$ & 
4&23(25)   & $C$     & 
3&92(21)   & $T$     & 
4&14(28)   & $C$     &
4&35(28)   & $T$     
\\
$\Phi_4$ & 
4&94(53)   & $C$     & 
4&06(28)   & $C$     & 
3&94(33)   & $T$     &
5&05(44)   & $C$  
\\
\hline
\end{tabular}
\caption{ \label{tab_02_0nf} 
  {\em Mass estimates and dominant operator contributions in the
    $0^{++}_{+}$ and $2^{++}_{+}$ channels for
    $N_f = 0$. The dominant operator types contributing are denoted
    with (without) parentheses if 
    $\langle \Phi_i^\dagger \phi_k\rangle <(>)0.5$.}}
\end{center}
\end{table}

\begin{table}[htb]
\begin{center}
\begin{tabular}{|l||r@{.}l|r@{.}l||r@{.}l|r@{.}l|r@{.}l||r@{.}l|}
\hline
&
\multicolumn{4}{c||}{$T = 1.5\lambdamsbar$} &
\multicolumn{8}{c|}{$T = 2.0\lambdamsbar$} \\
\cline{2-13}
&
\multicolumn{2}{c|}{$N_f = 2$} & 
\multicolumn{2}{c||}{$N_f = 3$} &
\multicolumn{2}{c|}{$N_f = 2$} & 
\multicolumn{2}{c|}{$N_f = 3$} & 
\multicolumn{2}{c||}{$N_f = 4$} & 
\multicolumn{2}{c|}{$N_f = 2$} \\
\cline{2-13}
&
\multicolumn{4}{c||}{$\beta=21$} &
\multicolumn{6}{c||}{$\beta=21$} & 
\multicolumn{2}{c|}{$\beta=28$} \\
\hline \hline
$0^{++}_+$  &  
\multicolumn{2}{c}{$M/g_3^2$} &
\multicolumn{2}{c||}{$M/g_3^2$} &   
\multicolumn{2}{c}{$M/g_3^2$} & 
\multicolumn{2}{c}{$M/g_3^2$} & 
\multicolumn{2}{c||}{$M/g_3^2$} & 
\multicolumn{2}{c|}{$M/g_3^2$} \\
\hline
$\Phi_1^{(R,L)}$ & 
1&183(22)   & 
1&316(18)   & 
1&351(18)   & 
1&477(18)   & 
1&557(22)   & 
1&335(19)   
\\
$\Phi_2^{(C)}$ & 
2&51(5)   & 
2&16(18)  & 
2&576(50) &
2&51(13)  & 
2&59(11)  & 
2&55(7)  
\\
$\Phi_3^{(T)}$ & 
2&64(9)   & 
2&86(19)  &
2&60(13)  &
2&90(18)  &
2&83(18)  &
2&74(9)  
\\
$\Phi_4^{(R,L)}$ & 
3&14(13)   &
3&26(18)   &
3&29(11)   &
3&35(21)   &
3&44(18)   &
3&25(12)  
\\
$\Phi_5^{(C)}$ & 
3&68(21)   &
3&83(24)   &
3&51(33)   &
3&86(25)   &
\multicolumn{2}{c||}{---}&
3&76(28)   
\\
$\Phi_6^{(T)}$ & 
4&01(35)   & 
3&84(32)   & 
4&11(25)   &
4&25(32)   &
\multicolumn{2}{c||}{---}&
4&05(33)  
\\
\hline\hline
$2^{++}_+$  &  
\multicolumn{2}{c}{$M/g_3^2$} & 
\multicolumn{2}{c||}{$M/g_3^2$} & 
\multicolumn{2}{c}{$M/g_3^2$} & 
\multicolumn{2}{c}{$M/g_3^2$} & 
\multicolumn{2}{c||}{$M/g_3^2$} & 
\multicolumn{2}{c|}{$M/g_3^2$}  \\
\hline
$\Phi_1^{(T)}$ & 
2&69(12)   & 
2&88(11)   & 
2&84(18)   &
2&98(12)   &
2&90(14)   &
2&74(7)  
\\
$\Phi_2^{(R,L)}$ & 
3&56(14)   &
3&73(15)   &
3&69(21)   &
3&57(28)   &
3&77(19)   &
3&59(12)   
\\
$\Phi_3^{(C)}$ & 
4&27(18)   &
4&08(28)   &
3&76(23)   &
4&27(33)   &
4&16(32)   &
4&14(24)  
\\
$\Phi_4^{(T)}$ & 
3&83(28)   &
4&40(33)   &
4&06(28)   &
\multicolumn{2}{c|}{---} &
\multicolumn{2}{c||}{---} &
3&87(24)   
\\
\hline
\end{tabular}
\caption{ \label{tab_02}
  {\em Mass estimates and dominant operator contributions in the
    $0^{++}_{+}$ and $2^{++}_{+}$ channels for various $N_f > 0$ on
    $L=30$, $40$ for $\beta =21$, $28$ respectively. The dominant
    operator types contributing are shown in the superscripts.}}
\end{center}
\end{table}

\begin{table}[htb]

\begin{center}
\begin{tabular}{|l||r@{.}l|r@{.}l||r@{.}l||r@{.}l|}
\hline
&
\multicolumn{6}{|c||}{$T=2T_c$} &
\multicolumn{2}{|c|}{$T \sim 10^{11}T_c$} \\
\cline{2-9}
&
\multicolumn{4}{|c||}{$\beta=21$}  &
\multicolumn{2}{|c||}{$\beta=28$} &
\multicolumn{2}{|c|}{$\beta=28$} \\
\cline{2-9}
&
\multicolumn{2}{|c|}{$L=38$} &
\multicolumn{2}{|c||}{$L=28$} &
\multicolumn{2}{|c||}{$L=40$} &
\multicolumn{2}{|c|}{$L=40$} \\
\hline\hline
$0^{--}_+$  & 
2&57 (11)   & 
2&53 (9)   & 
2&46 (10)   &
3&83 (19)   
\\
$0^{--*}_+$  & 
3&71 (28)   & 
3&91 (21)   & 
3&50 (25)   &
4&48 (28)   
\\
\hline
$0^{-+}_-$  & 
1&799 (35)   & 
1&771 (32)   & 
1&699 (43)   &
3&82 (12)   
\\
$0^{-+*}_-$  & 
3&28 (18)   & 
3&25 (18)   & 
3&27 (18)   &
5&07 (28)   
\\
\hline
$0^{+-}_-$  & 
1&85 (7)   & 
1&90 (5)   & 
1&85 (5)   &
4&96 (28)   
\\
$0^{+-*}_-$  & 
3&06 (33)   & 
3&06 (18)   & 
3&07 (12)  & 
\multicolumn{2}{c|}{---}
\\
\hline\hline
$1_+$ & 
3&47 (28) &
3&49 (30) &  
3&33 (28) &
\multicolumn{2}{c|}{---}
\\
\hline
$1_-$ & 
2&76 (28) &  
2&99 (21) &
2&98 (14) &
4&42 (28) 
\\
\hline
\end{tabular}
\end{center}

\caption{ \label{tab_other_0nf} 
{\em Masses $M/g_3^2$ for 
channels other than $0_+^{++}, 2_+^{++}$ at $N_f=0$.
The stars in the superscripts denote excited states. }}
\end{table}

\begin{table}[htb]

\begin{center}
\begin{tabular}{|l||r@{.}l|r@{.}l||r@{.}l|r@{.}l|r@{.}l||r@{.}l|}
\cline{1-13}
&
\multicolumn{4}{c||}{$T = 1.5\lambdamsbar$} &
\multicolumn{8}{c|}{$T = 2.0\lambdamsbar$} \\
\cline{2-13}
&
\multicolumn{2}{c|}{$N_f = 2$} & 
\multicolumn{2}{c||}{$N_f = 3$} &
\multicolumn{2}{c|}{$N_f = 2$} & 
\multicolumn{2}{c|}{$N_f = 3$} & 
\multicolumn{2}{c||}{$N_f = 4$} & 
\multicolumn{2}{c|}{$N_f = 2$} \\
\cline{2-13}
&
\multicolumn{4}{c||}{$\beta=21$} &
\multicolumn{6}{c||}{$\beta=21$} & 
\multicolumn{2}{c|}{$\beta=28$} \\
\hline\hline
$0^{--}_+$  & 
2&83 (7) & 
2&96 (7)  & 
2&74 (14) & 
2&84 (21) & 
3&16 (25) & 
2&83 (7)   
\\
$0^{--*}_+$  & 
3&71 (21) &  
3&94 (19) & 
3&70 (18) &  
3&58 (28) &  
3&84 (25) &  
3&64 (14)   
\\
\hline
$0^{-+}_-$  & 
1&932 (32) &  
1&977 (43)& 
1&960 (43) &  
2&079 (35) &  
2&103 (67) &  
1&932 (43)   
\\
$0^{-+*}_-$  & 
3&46 (11) &   
3&52 (21) & 
3&50 (21) &   
3&56 (11) &   
3&66 (16) &   
3&56 (14)   
\\
\hline
$0^{+-}_-$  & 
2&19 (7) &  
2&27 (11) & 
2&391 (35) &
2&604 (70) &
2&677 (70) &
\multicolumn{2}{c|}{---}
\\
$0^{+-*}_-$  & 
3&31 (24) &  
3&29 (19) & 
3&31 (15) &  
3&53 (21) &  
3&60 (11) &  
3&29 (10)  
\\
\hline
\hline
$1_+$ & 
3&59 (25) &
3&92 (20) & 
4&06 (16) &
4&11 (32) &
4&08 (25) &
3&73 (15) 
\\
\hline
$1_-$ & 
3&10 (18) &
3&17 (25) & 
2&91 (16) &
3&28 (28) &
3&21 (18) &
3&08 (12) 
\\
\hline
\end{tabular}
\end{center}

\caption{ \label{tab_other}
  {\em Masses $M/g_3^2$ for
    channels other than $0_+^{++}, 2_+^{++}$,
    for various $N_f > 0$,
    on $L=30$, $40$ with $\beta = 21$,~$28$, respectively.
    The stars in the superscripts denote excited states.}}
\end{table}

\begin{table}[htb]
\begin{center}
\begin{tabular}{|l||r@{.}l|r@{.}l||r@{.}l||r@{.}l|}
\hline
          &
\multicolumn{6}{|c||}{$T=2T_c$} &
\multicolumn{2}{|c|}{$T \sim 10^{11}T_c$} \\
\cline{2-9}
&
\multicolumn{4}{|c||}{$\beta=21$}  &
\multicolumn{2}{|c||}{$\beta=28$}  & 
\multicolumn{2}{|c|}{$\beta=28$} \\
\cline{2-9}
          &
\multicolumn{2}{|c|}{$L=38$} &
\multicolumn{2}{|c||}{$L=28$} &
\multicolumn{2}{|c||}{$L=40$} &
\multicolumn{2}{|c|}{$L=40$} \\
\hline
$aM_P(L)$  &
1&042 (40)   &
0&749 (20)   &
0&585 (17)   &
0&625 (20)   
\\
$a\sqrt{\sigma_{\infty}}$ &
0&167 (4) &
0&166 (3)   &
0&122 (2)   &
0&126 (2)   
\\
$\sqrt{\sigma_{\infty}}/g_3^2$ &
0&585 (15) &
0&580 (11)   &
0&569 (10)   &
0&588 (10)   
\\
\hline
\end{tabular}
\end{center}

\caption{ \label{tab_sigma_0nf}
{\em Spatial Polyakov loop masses and string tensions
for $N_f = 0$.}} 
\end{table}

\begin{table}[htb]

\begin{center}
\begin{tabular}{|l||r@{.}l|r@{.}l||r@{.}l|r@{.}l|r@{.}l||r@{.}l|}
\cline{1-13}
&
\multicolumn{4}{c||}{$T = 1.5\lambdamsbar$} &
\multicolumn{8}{c|}{$T = 2.0\lambdamsbar$} \\
\cline{2-13}
&
\multicolumn{2}{c|}{$N_f = 2$} &
\multicolumn{2}{c||}{$N_f = 3$} & 
\multicolumn{2}{c|}{$N_f = 2$} &
\multicolumn{2}{c|}{$N_f = 3$} &
\multicolumn{2}{c||}{$N_f = 4$} &
\multicolumn{2}{c|}{$N_f = 2$} \\
\cline{2-13}
&
\multicolumn{4}{c||}{$\beta=21$} & 
\multicolumn{6}{c||}{$\beta=21$} &
\multicolumn{2}{c|}{$\beta=28$} \\
\hline
$aM_P(L)$  &
0&760(21) &
0&818(15) &
0&798(19) &
0&847(35) &
0&820(19)&
0&588(11)
\\
$a\sqrt{\sigma_{\infty}}$ &
0&161(3)  &
0&167(2)  &
0&165(2)  &
0&170(4)  &
0&167(2)  &
0&123(2)
\\
$\sqrt{\sigma_{\infty}}/g_3^2$ &
0&563(11) &
0&584(8) &
0&577(8) &
0&595(15) &
0&584(8)&
0&574(10)
\\
\hline
\end{tabular}
\end{center}

\caption{ \label{tab_sigma}
{\em Spatial Polyakov loop masses and string tensions 
corresponding to 3d parameters with $N_f > 0$, 
on $L=30$, $40$ for $\beta=21$,~$28$ respectively.}}
\end{table}

\begin{table}[hbt]
\begin{center}
\begin{tabular}{|l||r@{.}l|r@{.}l|r@{.}l|r@{.}l|r@{.}l|}
\hline
 &  
\multicolumn{2}{c|}{$\mu/T = 0.5 \ii$}  &
\multicolumn{4}{c|}{$\mu/T = 1.0 \ii$}  & 
\multicolumn{4}{c|}{$\mu/T = 1.5 \ii$} \\
\cline{2-11}
 &  
\multicolumn{2}{c|}{$L=20$}  &
\multicolumn{2}{c|}{$L=20$}  &
\multicolumn{2}{c|}{$L=20$}  &
\multicolumn{2}{c|}{$L=30$}  &
\multicolumn{2}{c|}{$L=30$}  \\
 &  
\multicolumn{2}{c|}{rewtd.}  &
\multicolumn{2}{c|}{rewtd.}  &
\multicolumn{2}{c|}{exact}  &
\multicolumn{2}{c|}{rewtd.}  &
\multicolumn{2}{c|}{exact}  \\
\hline
\hline
$0^+$  &  
\multicolumn{2}{c|}{$M/g_3^2$} &
\multicolumn{2}{c|}{$M/g_3^2$} &
\multicolumn{2}{c|}{$M/g_3^2$} &
\multicolumn{2}{c|}{$M/g_3^2$} &
\multicolumn{2}{c|}{$M/g_3^2$}  \\
\hline
$\Phi_1^{(R_2,L)}$ &
1&34 (3)   & 
1&20 (2)   & 
1&18 (2)   & 
0&89 (3)  & 
0&85 (3)   \\
$\Phi_2^{(P)}$ &
1&94 (6) & 
1&89 (6) & 
1&73 (11) &
2&54 (14) & 
2&96 (21) \\
$\Phi_3^{(R_3)}$ &
2&21 (9) & 
2&22 (7) & 
2&21 (11) &
1&81 (6)  & 
1&81 (6)   \\
$\Phi_4^{(C)}$ &
2&54 (18) &
2&53 (9)  &
2&50 (13) &
2&45 (11)  & 
2&45 (18)   \\
\hline
\hline
$0^-$  &  
\multicolumn{2}{c|}{$M/g_3^2$} & 
\multicolumn{2}{c|}{$M/g_3^2$} & 
\multicolumn{2}{c|}{$M/g_3^2$} & 
\multicolumn{2}{c|}{$M/g_3^2$} & 
\multicolumn{2}{c|}{$M/g_3^2$} \\
\hline
$\Phi_1^{(B_1)}$ &
2&04 (3)   & 
1&87 (4)   & 
1&86 (3)   & 
1&76 (3)   & 
1&77 (7)   \\
$\Phi_2^{(B_2)}$ &
2&91 (15)  &
2&88 (9)   &
2&81 (13)  &
2&46 (7)   & 
2&42 (9)   \\
$\Phi_3^{(B_1)}$ &
3&15 (25)  & 
3&32 (18)  & 
3&45 (20)  & 
3&29 (21)  & 
3&23 (25)  \\
$\Phi_4^{(C)}$ &
3&80  (35) & 
3&75  (25) & 
3&63  (28) & 
3&50 (27)  & 
3&70 (28)  \\
\hline
\end{tabular}
\caption{ \label{tab_0_immu}
  {\em Mass estimates and dominant operator contributions in the $0^+$
    and $0^-$ channels for imaginary $\mu \neq 0$ from reweighted and
    exact actions. The dominant
    operator types contributing are shown in the superscripts.
    Some states on the lattice 
    size $L=30$ (used for $\mu/T=1.5i$)
    have been reordered for clarity of presentation.}}
\end{center}
\end{table}

\begin{table}[hbt]
\begin{center}
\begin{tabular}{|l||r@{.}lc|r@{.}lc|r@{.}lc|}
\hline
 &  
\multicolumn{3}{c|}{$\mu/T = 0.5$}  &
\multicolumn{3}{c|}{$\mu/T = 1.0$}  &
\multicolumn{3}{c|}{$\mu/T = 1.5$}  \\
\cline{2-10}
 &  
\multicolumn{3}{c|}{$L=30$}  &
\multicolumn{3}{c|}{$L=30$}  &
\multicolumn{3}{c|}{$L=30$}  \\
\hline
\hline
$0^+$  &  
\multicolumn{2}{c}{$M/g_3^2$} & Ops.  &
\multicolumn{2}{c}{$M/g_3^2$} & Ops.  &
\multicolumn{2}{c}{$M/g_3^2$} & Ops.  \\
\hline
$\Phi_1$ &
1&37 (4)   & $R_2,L$   & 
1&48 (2)   & $R_2,L$   &
1&59 (2)   & $R_2,L$   \\
$\Phi_2$ &
2&29 (8) & $C,(R_3)$   &
2&61 (3) & $C,(R_3)$   &
2&50 (5)   & $C$        \\
$\Phi_3$ &
2&51 (9) & $R_3,(C)$   &
2&61 (4) & $R_3,(C)$   &
2&78 (5)   & $R_3$       \\
$\Phi_4$ &
2&64 (17)  & $P$ &      
2&84 (8)  & $P$ &      
2&82 (7)  & $P$ \\
\hline
\hline
$0^-$  &  
\multicolumn{2}{c}{$M/g_3^2$} & Ops.  &
\multicolumn{2}{c}{$M/g_3^2$} & Ops.  &
\multicolumn{2}{c}{$M/g_3^2$} & Ops.  \\
\hline
$\Phi_1$ &
2&01 (2)   & $B_1$     &
2&02 (2)   & $B_1$     &
2&14 (3)   & $B_1$     \\
$\Phi_2$ &
2&84 (12)   & $B_2$   &
3&12 (6)   & $B_2$   &
3&18 (7)   & $B_2$   \\
$\Phi_3$ &
3&49 (18)   & $B_1$   &
3&56 (7)   & $B_1$   &
3&67 (11)  & $B_1$   \\
$\Phi_4$ &
3&74  (21) & $C$ & 
3&68  (11) & $C$ & 
3&73 (12)  & $C$ \\
\hline
\multicolumn{10}{c}{$\mbox{ }$} \\
\multicolumn{10}{c}{$\mbox{ }$} \\
\hline
 &  
\multicolumn{6}{c|}{$\mu/T = 2.0$}  &
\multicolumn{3}{c|}{$\mu/T = 4.0$}  \\
\cline{2-10}
 &  
\multicolumn{3}{c|}{$L=30$}  &
\multicolumn{3}{c|}{$L=40$}  &
\multicolumn{3}{c|}{$L=30$}  \\
\hline
\hline
$0^+$  &  
\multicolumn{2}{c}{$M/g_3^2$} & Ops.  &
\multicolumn{2}{c}{$M/g_3^2$} & Ops.  &
\multicolumn{2}{c}{$M/g_3^2$} & Ops.  \\
\hline
$\Phi_1$ &
1&73 (3)  & $R_2,L$   &
1&80 (11)  & $R_2,L$   &
2&58 (16) & $R_2,L$ \\
$\Phi_2$ &
2&45 (8)  & $C$       &
2&43 (21)  & $C$       &
2&69 (18) & $C$ \\
$\Phi_3$ &
3&01 (14)  & $R_3$        &
2&71 (21)  & $R_3$        &
\multicolumn{3}{c|}{---}  \\
$\Phi_4$ &
3&02 (14) & $P$ & 
\multicolumn{3}{c|}{---} &
\multicolumn{3}{c|}{---} \\
\hline
\hline
$0^-$  &  
\multicolumn{2}{c}{$M/g_3^2$} & Ops.  &
\multicolumn{2}{c}{$M/g_3^2$} & Ops.  &
\multicolumn{2}{c}{$M/g_3^2$} & Ops.  \\
\hline
$\Phi_1$ &
2&25 (5)   & $B_1$     &
2&19 (7)   & $B_1$     &
2&43 (20)  & $B_1$    \\
$\Phi_2$ &
3&45 (8)   & $B_2$   &
3&37 (25)   & $B_2$   &
4&17 (22)  & $B_2,C$   \\
$\Phi_3$ &
3&57 (11)  & $C$   &
3&78 (35)  & $C$   &
\multicolumn{3}{c|}{---} \\           
$\Phi_4$ &
3&70 (16)  & $B_1$ & 
3&85 (36)  & $B_1$ & 
\multicolumn{3}{c|}{---} \\
\hline
\end{tabular}
\caption{ \label{tab_0_remu}
  {\em Mass estimates and dominant operator contributions in the
    $0^+$ and $0^-$ channels for real
    $\mu \neq 0$. The dominant operator types contributing are denoted
    with (without) parentheses if 
    $\mid\langle \Phi_i^\dagger \phi_k\rangle\mid <(>)0.5$.}}
\end{center}
\end{table}

\end{document}